\documentstyle[12pt,aasms]{article}
\tightenlines

\begin{document}

ApJ Supplement, in press

\vskip 48pt

\title
{A Survey of Near Infrared Emission in Visual Reflection Nebulae}

\vskip 48pt

\author
{K. Sellgren$^ {1,2}$}

\vskip 12pt

\affil
{Department of Astronomy, Ohio State University,
174 West 18th Av., Columbus, OH  43210  USA}

\affil
{e-mail: sellgren@payne.mps.ohio-state.edu}

\vskip 24pt

\author
{M. W. Werner$^ 1$}

\vskip 12pt

\affil
{M/S 169-327, Jet Propulsion Laboratory, California
Institute of Technology, Pasadena, CA  91109}

\affil
{e-mail: mww@ipac.caltech.edu}

\vskip 24pt

\author
{L. J. Allamandola$^ 1$}

\vskip 12pt

\affil
{M/S 245-6, NASA Ames Research Center, Moffett Field, CA  94035}

\altaffiltext{1}
{Visiting astronomer at the Infrared Telescope Facility,
operated by the University of Hawaii under contract
with the National Aeronautics and Space Administration.}

\altaffiltext{2}
{Alfred P. Sloan Foundation Research Fellow}

\clearpage

\centerline
{\bf Abstract}

We present a survey for extended 2.2 $\mu$m emission in 20 new visual
reflection nebulae, illuminated by stars with temperatures of 3,600 --- 33,000
K.  We detect extended 2.2 $\mu$m emission in 13 new nebulae, illuminated by
stars with temperatures of 6,800 -- 33,000 K.  For most of these 13 nebulae we
have measured $J-K$, $H-K$, and $K-L'$, as well as obtaining surface brightness
measurements at the wavelength of the 3.3 $\mu$m emission feature.  All of the
reflection nebulae with extended near infrared emission in excess over
scattered starlight have very similar near infrared colors and show the
3.3 $\mu$m feature in emission with similar feature-to-continuum ratios.  The
3.3 $\mu$m feature-to-continuum ratio ranges from $\sim$3 to $\sim$9, both
within individual nebulae and from nebula to nebula, which suggests that the
3.3 $\mu$m feature and its underlying continuum arise from different materials,
or from different ranges of sizes within a size distribution of particles.  No
dependence on the temperature of the illuminating star is seen in the near
infrared colors or 3.3 $\mu$m feature-to-continuum ratio, over a factor of two
in stellar temperature.  This is similar to our previous IRAS results, in which
we found no dependence of the ratio of 12 $\mu$m to 100 $\mu$m surface
brightnesses in reflection nebulae illuminated by stars with temperatures of
5,000--33,000 K.


\clearpage

\section
{Introduction}

Extended near infrared continuum emission in visual reflection
nebulae was discovered by
Sellgren, Werner, \& Dinerstein (1983; hereafter Paper I)
and Sellgren (1984; hereafter Paper II) in NGC 2023, NGC 2068, and NGC 7023.
They showed that this near infrared continuum emission, unlike the visual
reflection nebulosity, was not primarily due to scattered starlight.
Sellgren, Werner, \& Dinerstein (1992; hereafter Paper IV) quantified
the amount of near infrared scattered starlight in NGC 2023 and NGC 7023
by polarization observations,
and showed that it contributes $<$20\% of the observed 2.2 $\mu$m
surface brightness of NGC 7023.
It was also shown in
Papers I and II
that thermal emission from
dust grains in equilibrium with the stellar radiation field is
unable to account for the near infrared continuum emission.
Observations of visual reflection nebulae by
Sellgren et al. (1985),
Castelaz, Sellgren, \& Werner (1987),
Sellgren, Luan, \& Werner (1990; hereafter Paper III),
and Roche, Aitken, \& Smith (1994) show that
this infrared continuum emission
extends at least from 1 to 25 $\mu$m.

A variety of proposals have been made to explain
this near infrared continuum emission.
One idea is
thermal emission from transiently heated tiny grains, with a
radius of $\sim$ 10 \AA, which are briefly heated to temperatures
near $\sim$ 1000 K by the absorption of single ultraviolet (UV) photons
(Papers I and II).
Another suggestion is
a quasi-continuum of overlapping
overtone and combination bands arising from
vibrational fluorescence in polycyclic
aromatic hydrocarbon (PAH) molecules
(L\'eger \& Puget 1984;
Allamandola, Tielens, \& Barker 1985).
A third explanation is
an electronic fluorescence in PAH molecules
(Allamandola, Tielens, \& Barker 1989).

The near infrared continuum emission in
the visual reflection nebulae NGC 2023, NGC 2068, and NGC 7023 is
strongly associated with an emission feature at 3.3 $\mu$m
(Paper I).
This emission feature has also been detected in
the visual reflection nebulae NGC 1333 (Whittet et al. 1983)
and Parsamyan 18 (Jourdain de Muizon, d'Hendecourt, \& Geballe 1990).
The 3.3 $\mu$m feature is one of a group of emission features at
3.3, 6.2, 7.7, 8.6, and 11.3 $\mu$m, observed together
in proto-planetary nebulae, planetary nebulae, reflection nebulae,
H II regions,
young stellar objects, and galaxies dominated by star formation
(see Aitken 1981, Willner 1984, Bregman 1989,
Puget \& L\'eger 1989, and Allamandola, Tielens, \& Barker 1989 for reviews).
The 6.2, 7.7, 8.6, and 11.3 $\mu$m features
have also been detected in the visual reflection nebulae
NGC 7023 (Sellgren et al. 1985),
NGC 2023 (Sellgren et al. 1985; Roche, Aitken, \& Smith 1994),
Parsamyan 18 (Cohen, Tielens, \& Allamandola 1985; Cohen et al. 1986),
NGC 2071 (Cohen et al. 1986),
and NGC 1333 (Roche, Aitken, \& Smith 1994).

The emission features seen at 3.3, 6.2, 7.7, 8.6, and 11.3 $\mu$m
are generally believed to be due to some sort of aromatic hydrocarbon.
Proposed laboratory analogs for the material which emits these
features include
amorphous carbon (Duley \& Williams 1981;
Borghesi, Bussoletti, \& Colangeli 1987),
C$_{60}$H$_{60}$ (Webster 1991),
coal (Papoular et al. 1989),
hydrogenated amorphous carbon (Blanco, Bussoletti, \& Colangeli 1988;
Ogmen \& Duley 1988),
nitrogenated amorphous carbon (Saperstein, Metin, \& Kaufman 1989),
the carbonaceous residue from Orgueil meteorite
(Wdowiak, Flickinger, \& Cronin 1988),
polycyclic aromatic hydrocarbons (L\'eger \& Puget 1984;
Allamandola, Tielens, \& Barker 1985, 1989;
L\'eger, d'Hendecourt, \& D\'efourneau 1989),
quenched carbonaceous composite (Sakata et al. 1987),
and a spiral carbonaceous microparticle
with an internal hydrogen (Balm \& Kroto 1990).
Most of these proposed laboratory analogs identify the 3.3 $\mu$m
feature with a C--H stretch vibration in aromatic hydrocarbons.
The emission mechanism for the aromatic hydrocarbon features
has been proposed to be
thermal emission from transiently heated tiny aromatic grains
(Papers I and II),
or a vibrational fluorescence in PAH molecules
(L\'eger \& Puget 1984; Allamandola, Tielens, \& Barker 1985, 1989).
Although similar materials and mechanisms have been
proposed in each case,
the emitting material and emission mechanism for the 3.3 $\mu$m feature and
the adjacent continuum may, or may not, be the same.

We present here the results of a survey of 20 additional
visual reflection nebulae for extended near infrared
continuum emission.
Our goal is to increase the sample of visual reflection
nebulae for which near infrared surface brightnesses and colors
are available, in order to see whether
the three bright nebulae observed in
Papers I and II are typical or extraordinary visual reflection nebulae.
We also present the results of a search for accompanying 3.3 $\mu$m emission
whenever extended near infrared continuum emission is detected.
We compare our observations to observations at
visual wavelengths, in order to quantify any possible contribution
from scattered starlight to the near infrared continuum emission.
We chose reflection nebulae with a range of values for
$T_{\rm star\ }$, the temperature of
the central illuminating star, to determine if there was
any dependence of the near infrared emission properties on $T_{\rm star\ }$.
Our sample was primarily chosen from the survey of visual reflection
nebulae conducted by van den Bergh (1966).
Table 1 gives a list of the nebulae we have observed,
and summarizes the properties of the central illuminating star
for each nebula.

\vskip 24pt

\section
{Observations}

The near infrared observations were primarily obtained at
the NASA Infrared Telescope Facility at Mauna Kea
Observatory,
between 1984 and 1988.
Observations were also obtained at the
United Kingdom Infrared Telescope
and at the University
of Hawaii 2.2-m telescope at Mauna Kea
Observatory
in 1984,
and at the 0.6-m telescope at Mount Wilson Observatory
in 1981.
The observational setup was the same as in Papers I and II.
Single detector InSb photometer systems cooled to solid
nitrogen temperatures were used in all cases.
For two nebulae, NGC 2071 and Parsamyan 18,
3.2--3.6 $\mu$m spectra were obtained with a
circular variable filter (CVF) having a spectral resolution
$\lambda$/$\Delta \lambda$ = 67, and
a 10.5$''$ diameter beam.
For the remaining sources studied at 3.3 $\mu$m,
however, we do not have complete spectra near the
feature, but only have
photometry at
the wavelength of the 3.3 $\mu$m feature,
obtained with CVFs
having spectral resolutions of $\lambda$/$\Delta \lambda$ =
50--100.
In these cases we estimated the
continuum near the 3.3 $\mu$m feature by interpolation
between $K$ (2.2 $\mu$m) and $L'$ (3.8 $\mu$m) surface photometry.

Chopping secondary mirrors were used for sky
subtraction, except for the 1981 observations
where the focal plane chopper described
by Becklin and Neugebauer (1968) was used.
The values used for the chopper throw
for each nebula are listed in Table 2.

The observational approach to this study of
extended near infrared emission in
reflection nebulae was to search first for emission
at $K$ at several nebular
positions, typically 42$''$ from the star.
Positions obviously contaminated by field stars
in the signal or reference beam were eliminated,
and if $K$ emission was detected in the
remaining positions, the brightest of these positions
was selected for further study.
Observations were obtained at this brightest position with a smaller aperture
to test whether the emission was truly extended,
then as much data as time permitted were obtained
at the brightest position
at $J$ (1.25 $\mu$m), $H$ (1.65 $\mu$m), $L'$,
and, using the CVF,
at the wavelength of the 3.3 $\mu$m emission feature.
For a few nebulae, multiaperture photometry at $K$ or
single aperture photometry at other infrared wavelengths
were obtained at more than one nebular position.

Because the observations entailed a search for
faint extended emission near bright stars,
careful measurements were made of the instrumental
scattered light.
Each offset position measured in a nebula was
also measured near a bright standard star, and the
scattered light correction determined for each
night applied to the observations of the
reflection nebulae.
At the standard offset positions given in Table 3{\it a},
the difference, $\Delta m$, between the magnitude of a star and the
magnitude of the instrumental scattered light correction measured
in a 10.5$''$ diameter aperture ranged from
$\Delta m$ = 8.7 mag to $\Delta m$ = 12.6 mag on different
nights and at different positions, with a
typical uncertainty in $\Delta m$
at a given position and night of 0.3 mag.
The resulting instrumental scattered light correction was
negligible for the brightest nebulae,
but was equal to or larger than the nebular emission
at some positions within
vdB 10, vdB 16, 23 Tau, Elias 1, vdB 34, vdB 35, vdB 46, vdB 101, vdB 111,
vdB 133, vdB 135, and IC 5076.
The uncertainties given for the observations therefore
include both the statistical and photometric uncertainties in the
individual observations, and the uncertainties in
the instrumental scattered light corrections.
For many nebulae in which no emission was detected,
particularly those with bright
illuminating stars, our observed upper limits are
dominated by the uncertainty in the instrumental scattered light
correction rather
than by the combination of statistical and photometric uncertainties.

Photometry was also obtained for the
central stars of many of the nebulae in Table 1,
at $J$, $H$, $K$,
$L'$, $N$ (10 $\mu$m), $Q$ (20 $\mu$m),
and, using the CVF,
at the wavelength of the 3.3 $\mu$m feature.
The $J$, $H$, $K$, and $L'$ stellar photometry was usually
obtained at the same time as the surface photometry of the
surrounding nebula, with the same observing setup,
as part of determining the instrumental scattered light correction.
The $N$ and $Q$ photometry was obtained at
the IRTF in 1987,
using a standard single-channel bolometer system cooled
with liquid helium, and a 6$''$ diameter aperture.

All observations were obtained on photometric nights.
The photometry and spectrophotometry were calibrated by observations of
stars from the IRTF standard star list (Tokunaga 1986),
with stars from
Elias et al. (1982) and the Caltech unpublished list preferred.
We assumed fluxes for zero magnitude of
1630 Jy, 1050 Jy, 655 Jy, 313 Jy, and 248 Jy at
$J$, $H$, $K$, 3.3 $\mu$m, and $L'$ respectively,
which were adopted from or interpolated between the
values of Cohen et al. (1992).
Scans of bright stars were used to correct the
photometry of extended sources for the
difference between a point source and uniform extended source.
This correction was not included in Paper I, and so we have applied
this correction to the results of Paper I when they are
given for comparison with the new results of the present paper.

\vskip 24pt

\section
{Results}

\subsection
{$K$ surface photometry}

The results of the $K$ surface
brightness photometry are given in Table 3.
Table 3{\it a} gives surface brightnesses measured at
standard offsets from the central star, of 30$''$E 30$''$N,
30$''$E 30$''$S, 30$''$W 30$''$N, and 30$''$W 30$''$S.
In Table 3{\it b} we present $K$ surface brightnesses
measured at other offsets,
generally chosen to complement data available at
other wavelengths.
We include in Table 3 results from Paper I for NGC 2023, NGC 2068, and NGC
7023,
corrected for an improved calibration, plus some new observations
of these sources.
Observations for 20 additional visual reflection nebulae are
presented here, which combined with the results
of Paper I provides $K$ surface brightnesses for 23
sources.

Of the 20 new nebulae observed, extended $K$ emission
was clearly detected in 13 of them.
Emission at $K$ at offsets away from the
central star was seen in two other
nebulae, vdB 135 and IC 5076, but observations
with different beam sizes were either unavailable or were unable to clearly
establish that the emission was extended.
No correction was attempted for possible
emission in the reference beam, which is
a potential problem for the most extended sources,
particularly 23 Tau.
For each nebula we combined all observations at the
same nebular offset, wavelength, and aperture size, without regard for
differences in the chopper amplitude or direction.

\subsection
{Nebular Colors}

In Table 4 we present the colors
of the reflection nebulae, at $J$, $H$, $K$, and $L'$.
We include for comparison the colors of
the three nebulae presented in Paper I, NGC 7023, NGC 2023,
and NGC 2068, again corrected for an improved calibration,
plus some new observations of these sources.
Some nebular colors were measured at multiple spatial positions,
or at the same spatial position but
with different aperture sizes.
Measurements of, or lower limits to, $J-K$ and $H-K$ were obtained for
all 15 nebulae in Table 4.
Measurements of, or upper limits on, $K-L'$
were obtained for 14 of the 15 nebulae in Table 4.

\subsection
{The 3.3 $\mu$m feature}

For two nebulae, NGC 2071 and Parsamyan 18,
3.2--3.6 $\mu$m spectra were obtained with a CVF.
Our spectrum of NGC 2071 is presented in Figure 1.
Our spectrum of Parsamyan 18, which is not shown, is similar to
the spectrum of this source published by
Jourdain de Muizon et al. (1990).
We did not have time for complete 3.2--3.6 $\mu$m spectra of
additional sources, so for the brightest reflection nebulae
photometry was obtained at the wavelength
of the 3.3 $\mu$m feature,
using the CVF.
These observations are included in Table 4,
in the form of a difference
between $K$ and the observed 3.3 $\mu$m
magnitude.
Measurements of, or upper limits on, $K-$[3.3 $\mu$m]
were obtained for 14 of the 15 nebulae in Table 4.

\subsection
{Photometry of the Central Stars}

In Table 5 we present the near-infrared magnitudes we have
measured for the central stars of the reflection nebulae
in this paper.
Magnitudes are given at $J$, $H$, $K$, and $L'$,
and at the wavelength of the 3.3 $\mu$m feature as measured through the
CVF.

Some central stars have an intrinsic 3.3 $\mu$m emission feature.
The spectrum of the pre-main sequence star Elias 1
shows emission features at 3.3, 3.43, and 3.53 $\mu$m
(Allen et al. 1982; Whittet et al. 1983;
Whittet, McFadzean, \& Geballe 1984; Schutte et al. 1990;
Tokunaga et al. 1991).
The 3.3 $\mu$m feature is detected in
$\sim$20\% of Herbig Ae/Be stars, and the 3.53 $\mu$m
feature is detected in $\sim$5\% of Herbig Ae/Be stars
(Brooke, Tokunaga, \& Strom 1993).
The 3.3 $\mu$m feature is not detected toward the
central stars of NGC 7023 (Paper II)
and NGC 1999 (Brooke et al. 1993), which are both
Herbig Ae/Be stars.

Other central stars
show a 3.3 $\mu$m emission feature due to nebular contamination,
because the aperture photometry for these stars contains
significant contributions from
the surrounding nebula included in the measurement.
This effect was seen in Paper II,
where the CVF spectrum of the central star of NGC 2023
showed the 3.3 $\mu$m feature in emission.
The central stars of NGC 1333 and Parsamyan 18, and probably
the central stars of NGC 2023, vdB 74, and NGC 7129,
have a measured magnitude at 3.3 $\mu$m
which is brighter than either the $K$ or $L'$ magnitude,
or which depends on aperture size.
The central stars of Parsamyan 18 and NGC 7129 are
Herbig Ae/Be stars without published 3 $\mu$m spectra,
but a comparison of their infrared magnitudes and the surface
brightness of the surrounding nebula indicates that any 3.3 $\mu$m
emission toward these stars would be most likely due to nebular
contamination rather than intrinsic emission from the star.

In Table 6 we present the mid-infrared magnitudes we have
measured for the central stars of the reflection nebulae
in this paper.
Magnitudes at 10 and 20 $\mu$m ($N$ and $Q$ respectively) are given.
We do not use these observations in our analysis but present
them for completeness.

\vskip 24pt

\section
{Discussion}

\subsection
{Energy Distributions}

In Paper II the observed broad-band energy distributions
from 1 to 1000 $\mu$m were presented for three reflection nebulae,
NGC 2023, NGC 2068, and NGC 7023.
These energy distributions show a strong near-infrared excess
over the expected level of near-infrared scattered starlight,
mid-infrared emission which is even stronger than the near-infrared
emission, and a peak in the far infrared presumably due to
thermal emission from grains in equilibrium with the radiation field.
In Paper III the IRAS energy
distributions from 12 to 100 $\mu$m were shown to be very similar for a
sample of reflection nebulae illuminated by stars with widely varying
$T_{\rm star\ }$.
In Figure 1 we plot the 0.44--3.8 $\mu$m broad-band energy distribution
of one of our observed reflection nebulae, NGC 2071,
along with its spectrum near the 3.3 $\mu$m feature.
Its visual and near infrared energy distribution
is very similar to that of previously observed reflection nebulae.
Its 3.2--3.6 $\mu$m spectrum is also very similar
to that of previously observed reflection nebulae (Paper I;
Whittet et al. 1983; Jourdain de Muizon et al. 1990).

For comparison with the observed nebular emission of NGC 2071,
we have estimated the scattered starlight for NGC 2071,
to show that while the visual emission is primarily scattered starlight,
the infrared emission is well in excess of the expected scattered starlight.
We predicted the amount of surface brightness, $S$, of scattered
starlight, relative to the stellar flux, $F_{\rm star\ }$,
using the model of Witt (1985$a$),
and assumed that all of the observed nebular emission at $V$ is
due to scattered starlight.
The prediction for the color of $S/F_{\rm star}$
between two wavelengths $\lambda_1$ and $\lambda_2$
in Equation 2 of Witt (1985$a$) is
$${{(S/F_{\rm star})_{\lambda_1}}\over{(S/F_{\rm star})_{\lambda_2}}} \ = \
{{\omega(\lambda_1) \ [1-e^{-\tau_0(\lambda_1)}]\
\exp[\tau_2(\lambda_1)-\tau_1(\lambda_1)]}
\over
{\omega(\lambda_2) \ [1-e^{-\tau_0(\lambda_2)}]\
\exp[\tau_2(\lambda_2)-\tau_1(\lambda_2)] \
p(\lambda_1,\lambda_2)}},\ \eqno(1)$$
where
$\tau_0(\lambda)$ is
the optical depth along the observed line of sight through the nebula,
$\tau_1(\lambda)$ is
the optical depth from the star to the nebular position observed,
$\tau_2(\lambda)$ is
the nebular optical depth from the star to the observer,
$\omega(\lambda)$ is the albedo,
and $p$ is a factor which depends on the nebular geometry and the wavelength
dependence of the phase function asymmetry.
We set $\tau_2$ = 0.5$\tau_0$, $\tau_1$ = 0, and $p$ = 1 for simplicity.
With these assumptions, the results are not sensitive to the adopted value
of $\tau_0$ as long as it is not much more than 1, the regime
in which the approximations made by Witt (1985$a$) break down.
We adopted the extinction law of Mathis (1990).
Theoretical and observational estimates of $\omega$($K$) range from
0.21 to $>$ 0.8
(Draine \& Lee 1984; Kim, Martin, \& Hendry 1994; Witt et al. 1994).
We have adopted two values for
$\omega$($\lambda$), a low albedo value (Draine \& Lee 1984)
and a high albedo value in which the albedo is independent of
wavelength from 0.44 to 3.8 $\mu$m.
We calculated the predicted scattered light with
both high and low values of the albedo.
Our results are shown in Figure 1, which demonstrates that at wavelengths
longer than 1 $\mu$m scattered starlight is not a significant contribution
to the observed emission of NGC 2071, for either assumption about the albedo.

\vskip 24pt
\subsection
{Nebular Colors}

The colors of the reflection nebulae are
all very similar, in close agreement with the
energy distribution of the nebulae whose observations are presented in
Paper I.
In Figure 2 we show a plot of $J-H$ vs.
$H-K$ for the nebulae,
to illustrate their similarity among nebulae.
All nebular positions for which we have $J-H$ and $H-K$ colors
are plotted.
Nebulae in which the 3.3 $\mu$m feature was detected are shown as
filled circles,
while nebulae in which the 3.3 $\mu$m feature was not searched
for, or searched for and not detected, are shown as
open circles.
We also show the colors of the central stars, with open stars for
normal stars and filled stars for stars with emission lines and thus
a probable infrared excess.

The reflection nebulae all have near infrared colors redder than their
illuminating stars, with the exceptions of
AE Aur, vdB 10, and NGC 1999.
These nebulae are the bluest nebulae in Figure 2.
The remaining nebulae in Figure 2 all have infrared colors
which are typical of the three reflection nebulae studied
in Papers I and II, and which characterize the energy distribution
of tiny grain and/or PAH continuum emission.
Thus the three reflection nebulae studied in Papers I and II,
chosen for their high surface brightness, are very representative
of the larger sample of reflection nebulae studied in the present paper.

A determination of the typical $V-K$ color of reflection nebulae
dominated by tiny grain and/or PAH emission is useful for estimating
the expected infrared surface brightness of reflection
nebulae for which only optical observations exist,
or the expected visual surface brightness of reflection
nebulae for which only infrared observations exist.
We have constructed $V-K$ colors for the nebulae for which
visual surface brightnesses were available at positions near
the positions observed in the near infrared
(Racine 1971;
Witt 1977;
Witt \& Cottrell 1980$a$;
Cottrell 1981;
Witt, Schild, \& Kraiman 1984;
Witt 1985$b$, 1986;
Witt \& Schild 1986;
Witt et al. 1987).
In Figure 3 we plot the $V-K$ color of the nebula, $(V-K)_{\rm neb\ }$, vs. the
observed $V-K$ color of the central star, $(V-K)_{\rm star\ }$.
As in Figure 2,
all nebular positions with $V-K$ colors are plotted,
with nebulae in which the 3.3 $\mu$m feature was detected shown as
filled circles,
and nebulae in which the 3.3 $\mu$m feature was not searched
for, or searched for and not detected, shown as
open circles.
Note that the observed stellar $V-K$ color is sometimes
considerably redder than expected for the observed $E(B-V)$ and
spectral type, because several stars in our
nebulae are Herbig Ae/Be stars with intrinsic infrared excesses
(Elias 1, V380 Ori, star A in Parsamyan 18, HD 200775, and BD +65 1637).
The nebulae with detected 3.3 $\mu$m emission features all have
$(V-K)_{\rm neb}$ $>$ 1.7, independent of $(V-K)_{{\rm star\ }}$.
NGC 1999 and 23 Tau have
significantly bluer $(V-K)_{\rm neb}$ colors compared to the nebulae
with detected 3.3 $\mu$m feature emission.

\subsection
{Scattered Starlight}

The fact that NGC 1999
has near infrared colors significantly bluer than its
illuminating star,
and that AE Aur and vdB 10
have near infrared colors similar to their
illuminating stars,
suggests the possibility that
in these particular sources the near infrared emission
could have a reflected starlight component.
In Paper IV the near infrared colors of the reflected starlight component
in NGC 7023 were
observed to be extremely blue compared to the colors of the illuminating
star, consistent with predictions for Rayleigh scattering
in an optically thin nebula: $\Delta (J-H)$ = $-1.21$ and
$\Delta (H-K)$ = $-1.25$,
where the difference is in the sense of the color of the nebula minus the
color of the star.
If we adopt a Mathis (1990) extinction law,
then Equation 1 predicts
$\Delta (J-H)$ = $-0.79$ and
$\Delta (H-K)$ = $-1.03$ for a low infrared albedo
(from Draine \& Lee 1984), and
$\Delta (J-H)$ = $-0.51$ and
$\Delta (H-K)$ = $-0.53$ for a high infrared albedo
(albedo independent of wavelength).
NGC 1999 is observed to have $\Delta (J-H)$ = $-0.77$ and
$\Delta (H-K)$ = $-0.84$,
colors which are
within the range of predictions for purely scattered starlight
in an optically thin nebula.

The relative amount of visual and near infrared nebular emission,
as measured by a color such as $V-K$, is another measure
of the relative importance of scattered
starlight and emission from tiny grains and/or PAH molecules.
The observed $V-K$ color of the central star, ($V-K$)$_{\rm star\ }$, depends
on the reddening of the star, which for many nebular geometries
is similar to the reddening of the nebula,
and on $T_{star}$.
The observed nebular $V-K$ color, ($V-K$)$_{\rm neb\ }$, depends
on ($V-K$)$_{\rm star\ }$,
on any differential reddening between nebula and star which is
usually much smaller than the total reddening,
and on the relative contributions of scattered
starlight and emission from tiny grains and/or PAH molecules to
the infrared emission.
Since the visual nebular emission is entirely due to scattered
starlight, the $V-K$ color should be smaller for
nebulae with infrared emission due only to scattered starlight,
and larger for nebulae with
both excess infrared emission and infrared scattered starlight.

We have drawn lines on Figure 3 for $(V-K)_{\rm neb}$
as a function of $(V-K)_{{\rm star}}$ for scattered starlight,
as predicted by Equation 1,
for four simple cases:
(1)
$\tau_0 (K)$/$\tau_0 (V)$ = 1
and $\omega (K)$/$\omega (V)$ = 1
(optically thick nebula, high albedo grains);
(2)
$\tau_0 (K)$/$\tau_0 (V)$ = 1
and $\omega (K)$/$\omega (V)$ = 0.38
(optically thick nebula, low albedo grains);
(3)
$\tau_0 (K)$/$\tau_0 (V)$ = 0.11
and $\omega (K)$/$\omega (V)$ = 1
(optically thin nebula, high albedo grains);
and (4)
$\tau_0 (K)$/$\tau_0 (V)$ = 0.11
and $\omega (K)$/$\omega (V)$ = 0.38
(optically thin nebula, low albedo grains).
For optically thin nebulae the value of $\tau_0 (K)$/$\tau_0 (V)$
is from Mathis (1990).
We predict $(V-K)_{\rm neb}-(V-K)_{\rm star}$
= 0.00, $-$1.06, $-$2.42, and $-$3.49,
respectively for Cases (1), (2), (3), and (4).
Nebulae with $K$ emission in excess of scattered starlight
fall above and to the left of these lines.

All of the nebulae with detected 3.3 $\mu$m emission in Figure 3
(shown as filled circles) fall above
and to the left of the lines for Cases (3) and (4),
and most fall above and to the left of the line for Case (2).
Note that the measurements of scattered starlight in NGC 7023
from the polarization results of Paper IV
fall between the lines for Case (2) and Case (4).
NGC 1999 also falls in this region, again suggesting its $K$
emission is primarily scattered starlight from an optically thin nebula.

Surface brightness measurements at $V$ are not available
for many of our reflection nebulae, particularly those
illuminated by cooler stars, and we did not obtain near infrared
colors for nebulae which we did not detect at $K$.
We therefore make an estimate of the surface
brightness of scattered starlight at $K$, $S_{\rm ref}$(est), in order to
compare our measurements of and upper limits on
the observed $K$ surface brightness, $S_{\rm obs\ }$, for our entire sample of
reflection nebulae.
The surface brightness of an optically thin reflection nebula
(Paper IV) is
$$S _ {\rm ref} \ = \ {\tau \ \omega \ F _ {\rm star}
\ H(g,\theta) \ \sin ^ 2 \theta
\over 4 \pi \ \phi ^ 2} \ . \eqno(2)$$
In this equation,
$\tau$ is the extinction optical depth,
$\omega$ is the grain albedo,
$F _ {\rm star}$ is the observed flux of the star,
and $\phi$ is the angular offset between star and nebula.
The star and nebula are assumed to be reddened by the same
amount in this equation.
The angular dependence of scattering is characterized by
the phase function
$H(g,\theta)$, where
$g$ = $< \cos \theta >$ is the phase asymmetry parameter
and $\theta$ is the scattering angle; the
functional form of $H(g, \theta)$ is given by
Henyey \& Greenstein (1941).
This surface brightness, when maximized with
respect to the scattering angle and phase function
(Sellgren 1983),
becomes
$$S _ {\rm ref} ({\rm est}) \ = \ {\tau \ \omega \ F _ {\rm star}
\over 4 \pi \ \phi ^ 2} \ . \eqno(3)$$
Thus $S _ {\rm obs} / S _ {\rm ref}({\rm est})$ $>$ 1 is a clear indication
of nebular emission from something other than reflected light.

We use two techniques to estimate a value of $\tau$($K$)
to use in Equation 3.
First we use the value of $A_V(neb)$ derived in Paper III,
which is an estimate of
$\tau$($V$) for each nebula calculated from the
amount of starlight incident on the nebula which is
reradiated in the far infrared.
Second we use Equation 3 combined with visual observations
of reflected light
(Racine 1971;
Witt 1977;
Witt \& Cottrell 1980$a$;
Cottrell 1981;
Witt, Schild, \& Kraiman 1984;
Witt 1985$b$, 1986;
Witt \& Schild 1986;
Witt et al. 1987)
to estimate $\tau$($V$) for each nebula.
The value of $\tau(V)$ was calculated from visual observations
at, or very near, each measured infrared position.
For IC 5076, we extrapolated data at offsets of
50$''$ and 100$''$ to a radius of 42$''$.
We then use a standard extinction curve (Mathis 1990)
to estimate $\tau$($K$) = $\tau$($V$) / 9.29.
We adopted a high infrared albedo, i.e.
assumed the albedo was independent of wavelength.
Our estimates of $\tau(K)$ are given in Table 7.

We plot our derived values of $S _ {\rm obs}$/$S _ {\rm ref}$(est)
vs. $T_{\rm star}$ in Figure 4.
All nebular positions with $K$ data and an estimate of $\tau$($V$) are shown.
If multiaperture
data were not able to demonstrate the $K$ emission is extended, that
detection was treated as an upper limit.
Nebulae in which the 3.3 $\mu$m feature was detected are shown in Figure 4 as
filled circles,
while nebulae in which the 3.3 $\mu$m feature was not searched
for, or searched for and not detected, are shown as
open circles.
Open triangles mark the reflected light measurements
in NGC 7023 (Paper IV).
We plot two independent points for nebulae
in which there were both IRAS estimates of
$A_V$(neb) from Paper III
and visual surface photometry of the nebula from which to estimate
$\tau$($V$).
These estimates agree to a factor of 3 on average, which provides
an estimate of our uncertainty in $\tau$($V$).

Figure 4 shows that most nebulae have an observed $K$
surface brightness which is above
the expected amount of reflected starlight we have estimated.
The only $K$ surface brightnesses low enough to
be consistent with purely scattered starlight are the measurements
of NGC 1999 and the upper limits on vdB 35, vdB 101, and vdB 135.
The observed surface brightness of reflected light in NGC 7023
(Paper IV)
is also consistent with the expected surface brightness
of reflected light, which is a check on our technique.
IC 5076 and Ced 201,
whose upper limits correspond to high $K$ surface brightnesses,
could have $K$ surface brightnesses which are lower than our upper limits,
and thus we cannot rule out these nebulae having $K$ surface
brightnesses which are also consistent with being due to scattered starlight.

We conclude that the $K$ extended emission
we detect is due to something other than reflected starlight
in most of our nebulae.
While the $K$ surface brightness of NGC 1999 is consistent with
predicted levels of scattered starlight,
{\it all} of the remaining sources detected at $K$ are strong
candidates for tiny grain and/or PAH emission at $K$.

\vskip 24pt
\subsection
{The 3.3 $\mu$m feature}

All of the sources where the 3.3 $\mu$m feature
was detected are characterized by $K$ surface
brightnesses which exceed those predicted for
reflected light (Figs. 1, 3, and 4), and by similar near infrared
energy distributions (Fig. 2), independent of $T_{\rm star\ }$.

Both the 3.3 $\mu$m feature-to-continuum ratio,
and the difference between $K$ and the magnitude at 3.3 $\mu$m,
show a similar range of values for all the sources in which
the 3.3 $\mu$m feature was detected.
This is illustrated in Figure 5,
which shows these two quantities
plotted vs. $T_{\rm star\ }$.
All nebular positions with measurements at these
wavelengths are shown.
The 3.3 $\mu$m feature-to-continuum ratio was derived by
interpolating a continuum between the $K$ and $L'$
continuum measurements, and so is not calculated for
nebulae for which there is no $L'$ detection.

Figure 5 shows that while the 3.3 $\mu$m feature-to-continuum
ratio has a similar range of values for all sources in which
the feature was detected, there is also a significant scatter
in this ratio, larger than the observational uncertainties.
The 3.3 $\mu$m feature-to-continuum
ratio varies from $\sim$3 to $\sim$9, with a median
of 5.3 and an average of 5.7.

It is possible
that some of the variation in the 3.3 $\mu$m feature-to-continuum
ratio is due to the $K$ emission being due to a mixture of tiny particle
emission and some other mechanism such as scattered starlight or
molecular hydrogen emission.
This would cause the continuum at 3.3 $\mu$m
to be overestimated, when interpolating between $K$ and $L'$,
and thus the 3.3 $\mu$m feature-to-continuum
ratio to be underestimated.
To investigate this further, we have
plotted in Figure 6 the 3.3 $\mu$m feature-to-continuum
ratio vs. $K-L'$.
All nebular positions with measurements at these
wavelengths are shown.
If the variation in the 3.3 $\mu$m feature-to-continuum
ratio is due to contamination of the $K$ emission by processes
other than the tiny particle emission which dominates the $L'$
emission,
then we would expect higher values of the 3.3 $\mu$m feature-to-continuum
ratio for larger values of $K-L'$.
No such trend is seen in Figure 6; there is no correlation observed
between the 3.3 $\mu$m feature-to-continuum
ratio and $K-L'$.

Another possible explanation is that the material which emits the
3.3 $\mu$m feature is not identical to the material
which emits the 3 $\mu$m continuum.
One example of this is the 3.3 $\mu$m feature
and the underlying continuum being due to
different ranges of a size distribution of aromatic particles,
but both groups of particles being excited by similar photon energies.
For instance, the feature might be due to vibrational fluorescence of
PAH molecules, and the continuum due
to electronic fluorescence of amorphous carbon grains or PAH clusters.
In this case, we would expect to see generally similar values of the
3.3 $\mu$m feature-to-continuum
ratio in regions of tiny particle emission,
as the particle size distribution is expected to be similar
throughout the interstellar medium,
but significant
variations in the ratio could be caused by slight modifications of
the particle size distribution by local conditions.
Such variations in the 3.3 $\mu$m feature-to-continuum
ratio might
be uncorrelated with any changes in the
mean photon energy ($T_{\rm star\ }$) or
shape of the continuum emission ($K-L'$),
as is seen in Figures 5 and 6.
The lack of correlation between 3.3 $\mu$m feature-to-continuum
ratio and $T_{\rm star}$ (Fig. 5), however, suggests that
the observed variation in the 3.3 $\mu$m feature-to-continuum
ratio is not due simply to differences in the required excitation energy
between the feature and its underlying continuum.

\vskip 24pt
\subsection
{The effect of the temperature of the illuminating star}

One goal of the observations presented in
this paper was
to test the proposals that the near infrared continuum emission
and 3.3 $\mu$m emission was due to thermal emission from tiny
grains stochastically heated by individual UV photons, as
proposed in Papers I and II, or due to vibrational
or electronic fluorescence from PAHs excited by UV radiation,
as proposed by L\'eger \& Puget (1984) and
Allamandola, Tielens, \& Barker (1985).
The original three visual reflection nebulae whose observations
are presented in Papers I and II
had central stars with $T_{\rm star}$ =
18,000--22,000 K.
Since the fraction of stellar luminosity available as UV photons is
a strong function of spectral type, we hoped by extending the
sample of visual reflection nebulae to hotter and cooler illuminating
stars that we would be able to test the UV excitation hypothesis directly.

In Paper III
IRAS surface photometry
of visual reflection nebulae was used to study how the tiny grain or
PAH emission depends on $T_{\rm star\ }$.
The ratio of surface brightness at 12 $\mu$m to
the surface brightness at 100 $\mu$m, $I_{12}/I_{100}$,
was used
as a measure of the relative amount of stellar energy absorbed and
re-radiated by tiny grains and/or PAHs (at 12 $\mu$m) and
by large grains in equilibrium with the stellar radiation field
(at 100 $\mu$m).
It was found that $I_{12}/I_{100}$ is independent of $T_{\rm star}$
for 5,000 K $<$ $T_{\rm star}$ $<$ 33,000 K.
This is inconsistent with excitation of the 12 $\mu$m emission
by UV radiation alone,
but is consistent with the 12 $\mu$m emission
being excited by photons with wavelengths ranging
from 0.1 to 0.7 $\mu$m.
Because PAHs absorb most strongly at wavelengths
$<$0.4 $\mu$m
(Crawford, Tielens, \& Allamandola 1985;
L\'eger et al. 1989;
Allamandola, Tielens, \& Barker 1989),
the results of Paper III suggest that
the 12 $\mu$m IRAS emission
is not due solely to free-flying PAH molecules.

The observation
that $I_{12}/I_{100}$ is independent of $T_{\rm star}$
(Paper III)
does not rule out some contribution by PAH molecules to the
12 $\mu$m emission, but does show that
there must be at least one component to the 12 $\mu$m emission
that absorbs over a wider range of UV and visual wavelengths
than the narrow range of wavelengths at which PAHs absorb.
The IRAS 12 $\mu$m band covers 8.0 -- 14.9 $\mu$m
(FWHM, from IRAS Explanatory Supplement)
and so can include emission from several components:
strong emission features at 7.7, 8.6, and 11.3 $\mu$m;
a weaker emission feature at 12.7 $\mu$m;
a plateau of emission at 11--13 $\mu$m; and
continuum emission throughout the filter
(see Allamandola, Tielens, \& Barker 1989 for a review).
Roche, Aitken, \& Smith (1989) and
Bregman et al. (1989) observe that the 11.3 $\mu$m emission
has a different spatial distribution in the Orion Bar than the nearby
continuum emission, indicating these arise from
distinct components of the interstellar medium.
Bregman et al. (1994) find that images at 3.3, 8.4, and 11.3 $\mu$m
of the Orion Bar all look different, again
supporting the idea that different components are responsible
for emission in different features.
Allamandola, Tielens, \& Barker (1989) have identified
emission from PAHs, PAH clusters, and amorphous carbon grains
as contributing to different emission components seen
in the IRAS 12 $\mu$m band.
Thus the results of Paper III could be consistent with PAH molecular
emission contributing to the 12 $\mu$m IRAS emission, as long as
there is also one or more other materials,
such as PAH clusters or amorphous carbon
grains, which are able to absorb over a wide wavelength range in the
visible and UV and emit within the 12 $\mu$m IRAS band.

If the extended near infrared emission, in excess
of reflected light, is
due to some sort of non-equilibrium thermal emission
from tiny grains (Paper II),
then it would be interesting to search for temperature
differences between nebulae with different values of $T_{\rm star\ }$.
The best indicator of tiny grain temperature should be $K-L'$,
because the excess infrared emission is stronger,
relative to any reflected light component, at
longer wavelengths (Sellgren et al. 1985; Papers III and IV).
If instead the emission is due to UV fluorescence
from PAHs (L\'eger and Puget 1984;
Allamandola et al. 1985)
then $K-L'$ is an indicator of whether the emitted fluorescent
spectrum changes with excitation.
Figure 7 shows $J-H$, $H-K$, and $K-L'$ plotted vs. $T_{\rm star\ }$.
All nebular positions with measurements at these
wavelengths are shown.
There is no evidence for changing $J-H$, $H-K$, or $K-L'$ color with
$T_{\rm star}$.
The nebular offsets for the near infrared data ($\sim$42$''$)
and the IRAS data from Paper III ($\sim$180$''$) were too different for
a comparison of the near and mid infrared data.

We do not see any dependence on $T_{\rm star}$
for either the 3.3 $\mu$m feature-to-continuum ratio
or the $K-L'$ color among the stars for which we have such
observations (Figs. 5 and 7),
over a factor of two range in $T_{\rm star\ }$.
In Paper III we found that
the nebulae illuminated by the coolest stars in our sample
provided the necessary range in $T_{\rm star}$ to rule out
the IRAS 12 $\mu$m emission being due purely to PAH molecules.
We would have liked to extend our observations of
the 3.3 $\mu$m feature-to-continuum ratio
or the $K-L'$ color to even lower
values of $T_{\rm star\ }$, but we found that
the visual reflection nebulae with higher $K$ surface brightnesses
were largely confined to the hottest illuminating stars in our sample.
We did not have enough sensitivity
to obtain near infrared colors and 3.3 $\mu$m spectrophotometry
in visual reflection nebulae with $K$ surface brightnesses
below 0.3 MJy sr$^{-1}$.
Thus our observations currently do not have enough range in $T_{\rm star}$
to confirm or rule out models in which the near infrared continuum emission or
3.3 $\mu$m emission
is due to PAH molecules.
This test must await more sensitive observations, such as may be
possible from ISO, SIRTF, or IRTS.

\section
{Summary}

We have extended the observations of Papers I and II, to
a survey of near infrared emission in a total of 23 visual reflection nebulae.
Extended near infrared continuum emission at 1--5 $\mu$m and
emission from a spectral feature at 3.3 $\mu$m
in the three visual reflection nebulae studied in Papers I and II
has been attributed to either thermal emission from
stochastically heated tiny grains (Papers I and II) or
to vibrational or electronic fluorescence
from polycyclic aromatic hydrocarbon (PAH) molecules
(L\'eger \& Puget 1984; Allamandola, Tielens, \& Barker 1985, 1989).
Our new observations report on a search
for extended 2.2 $\mu$m emission in
visual reflection nebulae
illuminated by stars with spectral types ranging from O9.5V to M1IIIe.

We have detected extended 2.2 $\mu$m emission,
after correction for instrumental scattered light,
in 16 out of 23 nebulae studied,
illuminated by stars with spectral types ranging from O9.5V to F5Iab.
For most of these 16 nebulae we have measured $J-K$, $H-K$, and $K-L'$,
as well as obtaining
surface brightness measurements at the wavelength of
the 3.3 $\mu$m emission feature.
We use several approaches to predict the $K$ surface brightness
of scattered starlight in these nebulae, in order to determine whether our
$K$ surface brightnesses are consistent with, or in excess over,
scattered starlight.
The only source which has near infrared colors and a $K$ surface
brightness consistent with being due primarily to scattered starlight
is NGC 1999; the remaining sources with extended $K$ emission
are all consistent with some or all of their infrared emission being due
to tiny grains and/or PAHs.

All of the reflection nebulae with extended near infrared emission
in excess over scattered starlight
have very similar near infrared colors and show the 3.3 $\mu$m feature
in emission with similar feature-to-continuum ratios.
The three reflection nebulae previously studied in Papers I and II,
while having high surface brightnesses,
are still very representative of our larger sample.
The 3.3 $\mu$m feature-to-continuum
ratio ranges from $\sim$3 to $\sim$9, both within individual
nebulae and from nebula to nebula;
this variation does not appear to depend on $T_{\rm star}$ or $K-L'$.
One possible explanation for this is that the 3.3 $\mu$m feature
and its underlying continuum arise from different materials, or
from different ranges of sizes within a size distribution of
particles.
No dependence on $T_{\rm star}$ is seen
in the near infrared colors or 3.3 $\mu$m feature-to-continuum ratio,
over a factor of two range in $T_{\rm star\ }$.
We do not have the sensitivity at present
to obtain a wider range in $T_{\rm star\ }$,
to test whether or not this is in
conflict with models of tiny grain or PAH emission.
ISO, SIRTF, or IRTS
may be able to make these tests with higher sensitivity in future.

\vskip 24pt

\acknowledgments

We would like to thank the IRTF and UKIRT staffs for
their superb technical support.
We also thank Tom Geballe, Bill Golisch, Dave Griep, Bob Howell,
Charlie Kaminski, Martina McGinn, Mark Shure, Xander Tielens, Alan Tokunaga,
and Doug Toomey for assistance with the observations.
We are grateful to Darren DePoy, Jay Frogel, and Doug Whittet
for suggestions on improving our paper.
This work was carried out in part at the Jet Propulsion Laboratory,
California Institute of Technology,
under contract with the National Aeronautics and Space Administration.
This research has made use of the Simbad database,
operated at CDS, Strasbourg, France.

\clearpage

\begin{center}
\begin{tabular}{rllcclcl}
\multicolumn{8}{c}{{\bf TABLE 1}}\\[12pt]
\multicolumn{8}{c}{{\bf Characteristics of Nebular Illuminating Stars}}\\[12pt]
\multicolumn{1}{r}{vdB}&\multicolumn{1}{l}{Nebula}&\multicolumn{1}{l}{Star}
&&&\multicolumn{1}{l}{Spectral}&\multicolumn{1}{c}{$T _ {\rm star}$}\\
\multicolumn{1}{r}{No.$^ a$}&\multicolumn{1}{l}{Name}
&\multicolumn{1}{l}{Name}&\multicolumn{1}{c}{$V$}
&\multicolumn{1}{c}{$E(B-V)$}&\multicolumn{1}{l}{Type}
&\multicolumn{1}{c}{(K)$^c$}&\multicolumn{1}{l}{Ref.$^ d$}\\[12pt]
10&...&HD 20041&5.81&0.71&A0Ia&9,400&1\\
16&...&BD+29 565&9.16&0.30&F0V&7,000&1\\
17&NGC 1333&BD+30 549&10.47&0.60&B8V&11,000&1\\
22&Merope Nebula$^ b$&23 Tau$^ b$&4.18&0.09&B6IVnn&12,000&1\\
$$...&IC 359&Elias 1$^b$&16.2&1.3$^e$&A6e&8,100&2, 3\\
34&...&AE Aur$^ b$&5.97&0.53&O9.5V&33,000&1\\
35&...&HD 34033&8.66&0.13&G8II--III&4,900&1\\
46&NGC 1999&V380 Ori$^ b$&10.22&0.70&A0--2e&9,500&1, 4\\
52&NGC 2023&HD 37903&7.82&0.36&B1.5V&22,000&1\\
57&IC 435&HD 38087&8.30&0.35&B3n&18,000&1\\
59&NGC 2068&HD 38563N&10.56&1.42&B2II--III&19,000&5\\
60&NGC 2071&HD 290861a&10.14&1.25&B2--3&19,000&5\\
72&...&HD 42261&9.18&0.41&B3V&18,000&1\\
74&...&BD$-$6 1444&10.82&0.63&B6&13,000&1\\
$$...&Parsamyan 18$^ b$&star A&13.21&1.26&B2--3e&19,000&6\\
101&...&HD 146834&6.35&0.33&G5III&5,000&1\\
111&...&HD 156697&6.50&0.11$^f$&F0n&7,000&1\\
133&...&HD 195593&6.19&0.65&F5Iab&6,800&1\\
135&...&BD+31 4152&8.43&0.40&M1IIIe&3,600&1\\
137&IC 5076&HD 199478&5.69&0.50&B8Ia&10,000&1\\
139&NGC 7023&HD 200775&7.39&0.44&B3e&18,000&1, 7\\
146&NGC 7129&BD+65 1637&10.15&0.67&B2nne&21,000&1\\
152&Ced 201&BD+69 1231&9.29&0.21&B9.5V&10,000&1\\[24pt]
\end{tabular}
\end{center}

\noindent
($a$) Number in reflection nebula catalog of
van den Bergh (1966).

\noindent
($b$) Other designations: Merope nebula = NGC 1435;
23 Tau = HD 23480;
Elias 1 = V892 Tau;
AE Aur = HD 34078;
V380 Ori = BD$-$6 1253;
Parsamyan 18 = NGC 2316.

\noindent
($c$) Temperature of central star, $T_{\rm star}$,
derived from spectral type using calibrations
of Panagia (1973) for B3 and earlier stars and
Johnson (1966) for the remaining stars.

\noindent
($d$) References---(1) Racine 1968;
(2) Elias 1978;
(3) Cohen \& Kuhi 1979;
(4) Herbig 1960;
(5) Strom et al. 1975;
(6) see note ($g$) to this Table;
(7) Witt \& Cottrell 1980$b$.

\noindent
($e$)
$E(B-V)$ for Elias 1 derived from $B-V$ = 1.5 (Elias 1978)
and adopting $(B-V)_0$ = 0.17 for A6V (Johnson 1966).

\noindent
($f$)
$E(B-V)$ for HD 156697 derived from $B-V$ = 0.41 (Racine 1968)
and adopting $(B-V)_0$ = 0.31 for F0V (Johnson 1966).

\noindent
($g$)
Parsamyan 18 contains a B2--3 star which excites
a compact H II region (L\'opez et al. 1988).
L\'opez et al. (1988) suggest that the two condensations visible
in Parsamyan 18 are not stars but rather
knots of nebulosity due to light reflected
from an obscured H II region.
These two condensations, however, have stellar spectra and are
not identical;
the northeast component, star A, is a B emission line star,
and the southwest component is an A--F star
(Humphreys 1985; Strom 1985).
It seems reasonable to identify star A with the exciting star of the
H II region, giving a spectral type of B2--3e.
Star A has $V$ = 13.21 and $B-V$ = 1.04 (Humphreys 1985), which
was combined with an assumed $(B-V)_0$ = $-0.22$ for B2.5V (Johnson 1966)
to derive $E(B-V)$.

\clearpage
\begin{center}
\begin{tabular}{ll}
\multicolumn{2}{c}{{\bf TABLE 2}}\\[12pt]
\multicolumn{2}{c}{{\bf Chopper Throws Used}}\\[12pt]

Name  & Chopper Throws \\

vdB 10 &   180$''$ EW\\

vdB 16 &   180$''$ EW\\

vdB 17 &    120$''$ EW, 180$''$ NS\\

vdB 22 &    180$''$ EW, 180$''$ NS, 342$''$ EW, 377$''$ EW\\

Elias 1 &   180$''$ EW, 180$''$ NS\\

vdB 34 &     120$''$ EW, 180$''$ EW\\

vdB 35 &   180$''$ EW\\

vdB 46 &   180$''$ EW\\

vdB 52 &     210$''$ EW, 250$''$ EW\\

vdB 57 &   180$''$ EW, 180$''$ NS\\

vdB 59 &     180$''$ EW, 210$''$ EW\\

  vdb 60 &   180$''$ EW, 180$''$ NS\\

  vdb 72 &   180$''$ EW\\

  vdb 74 &   180$''$ EW\\

Parsamyan 18 &    120$''$ EW, 180$''$ EW\\

vdb 101 &   180$''$ EW\\

vdb 111 &   180$''$ EW\\

vdb 133 &     120$''$ EW, 180$''$ EW, 180$''$ NS\\

vdb 135 &   180$''$ EW\\

vdb 137 &   120$''$ EW, 180$''$ EW\\

vdb 139 &   120$''$ EW, 150$''$ EW, I 160$''$ EW\\

 vdb 146 &     180$''$ EW, 180$''$ NS\\

 vdb 152 &   180$''$ NS\\

\end{tabular}
\end{center}
\clearpage

\begin{center}
\begin{tabular}{lrrrrr}
\multicolumn{6}{c}{{\bf TABLE 3a}}\\[12pt]
\multicolumn{6}{c}{{\bf 2.2 $\mu$m Surface Brightness of Reflection
Nebulae}}\\[12pt]
&&\multicolumn{4}{c}{$S _ \nu$ (MJy sr$ ^ {-1}$) at Given Nebular
Offset$^b$}\\[12pt]
Name&Ap$^a$&30$''$E 30$''$N&30$''$E 30$''$S& 30$''$W 30$''$N&30$''$W
30$''$S\\[12pt]
vdB   10 &  6.2$''$&...&...&   2.02 &...\\
&& & &$\pm$  0.12 & \\
 &  7.7$''$&   2.15 &...&   1.73 & $<$  0.42 \\
&&$\pm$  0.27 & &$\pm$  0.20 & \\
 & 10.5$''$&   0.23 &...&   1.64 &...\\
&&$\pm$  0.07 & &$\pm$  0.12 & \\
vdB   16 & 10.5$''$& $<$  0.16 & $<$  0.10 & $<$  0.12 & $<$  0.08 \\
&& & & & \\
vdB   17 &  8.2$''$& $<$  1.36 &   3.26 &...& $<$  2.89 \\
&& &$\pm$  0.87 & & \\
vdB   22 & 10.5$''$& $<$  2.75 & $<$  0.27 & $<$  0.37 &   0.31 \\
&& & & &$\pm$  0.05 \\
 & 19.6$''$&...&...&...&   0.35 \\
&& & & &$\pm$  0.09 \\
Elias 1 &  6.2$''$& $<$  0.69 &...&...&...\\
&& & & & \\
 & 10.5$''$& $<$  0.68 &...& $<$  0.08 & $<$  0.13 \\
&& & & & \\
vdB   34 &  6.2$''$&...&   0.27 &...&...\\
&& &$\pm$  0.05 & & \\
 & 10.5$''$&   0.47 &   0.51 & $<$  0.55 &...\\
&&$\pm$  0.06 &$\pm$  0.05 & & \\
 & 19.6$''$&...&   0.34 &...&...\\
&& &$\pm$  0.04 & & \\
vdB   35 & 10.5$''$& $<$  0.51 & $<$  0.06 & $<$  0.10 & $<$  0.11 \\
&& & & & \\
vdB   46 &  6.2$''$&...&...&   0.45 &...\\
&& & &$\pm$  0.13 & \\
 & 10.5$''$& $<$  0.46 &   0.29 &   0.45 &   0.21 \\
&& &$\pm$  0.03 &$\pm$  0.05 &$\pm$  0.04 \\
\end{tabular}
\end{center}
\begin{center}
\begin{tabular}{lrrrrr}
\multicolumn{6}{c}{{\bf TABLE 3a, Continued}}\\[12pt]
\multicolumn{6}{c}{{\bf 2.2 $\mu$m Surface Brightness of Reflection
Nebulae}}\\[12pt]
&&\multicolumn{4}{c}{$S _ \nu$ (MJy sr$ ^ {-1}$) at Given Nebular
Offset$^b$}\\[12pt]
Name&Ap$^a$&30$''$E 30$''$N&30$''$E 30$''$S& 30$''$W 30$''$N&30$''$W
30$''$S\\[12pt]
vdB   57 &  6.2$''$&...&   0.33 &...&...\\
&& &$\pm$  0.07 & & \\
 & 10.5$''$&...&   0.43 &...&...\\
&& &$\pm$  0.05 & & \\
vdB   60 &  6.2$''$&...&   6.14 &...&...\\
&& &$\pm$  0.35 & & \\
 & 10.5$''$&   5.09 &   5.52 &   1.41 &   5.13 \\
&&$\pm$  0.34 &$\pm$  0.31 &$\pm$  0.08 &$\pm$  0.29 \\
vdB   72 &  6.2$''$&...&...&...&   1.84 \\
&& & & &$\pm$  0.14 \\
 & 10.5$''$&...&...&...&   1.78 \\
&& & & & \\
vdB   74 &  6.2$''$&   0.67 &...&...&...\\
&&$\pm$  0.08 & & & \\
 & 10.5$''$&   0.57 &   0.54 &...&   0.47 \\
&&$\pm$  0.05 &$\pm$  0.09 & &$\pm$  0.09 \\
 & 19.6$''$&   0.47 &...&...&...\\
&&$\pm$  0.07 & & & \\
vdB  101 &  7.7$''$& $<$  0.32 & $<$  0.15 & $<$  0.13 & $<$  0.18 \\
&& & & & \\
vdB  111 &  6.2$''$&...&   0.22 &...&...\\
&& &$\pm$  0.03 & & \\
 &  7.7$''$&...&   0.22 &...& $<$  0.03 \\
&& &$\pm$  0.02 & & \\
\end{tabular}
\end{center}
\begin{center}
\begin{tabular}{lrrrrr}
\multicolumn{6}{c}{{\bf TABLE 3a, Continued}}\\[12pt]
\multicolumn{6}{c}{{\bf 2.2 $\mu$m Surface Brightness of Reflection
Nebulae}}\\[12pt]
&&\multicolumn{4}{c}{$S _ \nu$ (MJy sr$ ^ {-1}$) at Given Nebular
Offset$^b$}\\[12pt]
Name&Ap$^a$&30$''$E 30$''$N&30$''$E 30$''$S& 30$''$W 30$''$N&30$''$W
30$''$S\\[12pt]
vdB  133 &  6.2$''$& $<$  4.71 &...&...&...\\
&& & & & \\
 &  8.2$''$& $<$  2.49 & $<$  3.21 & $<$  3.36 & $<$  3.10 \\
&& & & & \\
 & 10.5$''$&   2.14 & $<$  0.24 & $<$  0.28 & $<$  0.25 \\
&& & & & \\
 & 19.6$''$&   2.82 &...&...&...\\
&&$\pm$  0.35 & & & \\
vdB  135 &  7.7$''$& $<$  0.37 & $<$  0.16 &...&   0.65 \\
&& & & &$\pm$  0.09 \\
vdB  137 &  6.2$''$&...&...&...& $<$  0.28 \\
&& & & & \\
 & 10.5$''$& $<$  0.12 &...& $<$  0.14 &   0.39 \\
&& & & &$\pm$  0.09 \\
vdB  146 &  6.2$''$&   1.36 &...&...&...\\
&&$\pm$  0.13 & & & \\
 &  8.2$''$& $<$  2.66 &   3.14 &...& $<$  1.41 \\
&& &$\pm$  0.60 & & \\
 &  9.2$''$&   0.92 &...&...&...\\
&& & & & \\
 & 10.5$''$&   1.14 &   5.24 &   2.08 &...\\
&&$\pm$  0.09 &$\pm$  0.32 &$\pm$  0.12 & \\
vdB  152 &  8.2$''$& $<$  1.96 & $<$  2.15 & $<$  1.47 & $<$  0.79 \\
&& & & & \\
\end{tabular}
\end{center}
\clearpage
\begin{center}
\begin{tabular}{lrrrrr}
\multicolumn{6}{c}{{\bf TABLE 3b}}\\[12pt]
\multicolumn{6}{c}{{\bf 2.2 $\mu$m Surface Brightness of Reflection
Nebulae}}\\[12pt]
&&\multicolumn{2}{c}{Nebular Offset$^b$}&$S _ \nu$ (2.2 $\mu$m)\\
Name&Ap$^a$&$\Delta \alpha$&$\Delta \delta$&(MJy sr$ ^ {-1}$)\\[12pt]
vdB   17 &  6.2$''$&   0$''$E&  20$''$S&   1.12 &$\pm$  0.14 \\
 &  8.2$''$&  30$''$W&   0$''$N& $<$  2.72 \\
 &  8.2$''$&  30$''$E&   0$''$N& $<$  3.94 \\
 & 10.5$''$&   0$''$E&  20$''$S&   1.24 &$\pm$  0.07 \\
vdB   22 &  8.2$''$&  30$''$W&   0$''$N& $<$  2.73 \\
 &  8.2$''$&  30$''$E&   0$''$N& $<$  3.24 \\
 & 10.5$''$&   0$''$E& 557$''$S& $<$  0.27 \\
 & 10.5$''$&   0$''$E& 135$''$S& $<$  0.09 \\
 & 10.5$''$&   0$''$E& 125$''$S&   0.09 &$\pm$  0.03 \\
Elias 1 &  8.2$''$&  30$''$W&   0$''$N& $<$  3.12 \\
 &  8.2$''$&  30$''$E&   0$''$N& $<$  4.31 \\
vdB   52 & 10.5$''$&  40$''$W&  40$''$S&   5.23 &$\pm$  0.73 \\
 & 10.5$''$&   0$''$E&  60$''$S&   6.59 &$\pm$  0.91 \\
 & 10.5$''$&   0$''$E&  60$''$N&   2.81 &$\pm$  0.32 \\
vdB   59 & 10.5$''$&  60$''$E&  40$''$S&   1.57 &$\pm$  0.09 \\
vdB   60 &  8.2$''$&  30$''$E&   0$''$N&   4.47 &$\pm$  1.03 \\
Parsamyan 18 &  6.2$''$&  12$''$W&   0$''$N&   7.13 &$\pm$  0.92 \\
 &  6.2$''$&  12$''$W&  12$''$N&   3.28 &$\pm$  0.35 \\
 &  6.2$''$&   0$''$E&  12$''$S&   7.46 &$\pm$  0.81 \\
 &  6.2$''$&   0$''$E&  12$''$N&   6.04 &$\pm$  0.67 \\
 &  6.2$''$&  12$''$E&   0$''$N&   2.19 &$\pm$  0.36 \\
 & 10.5$''$&  15$''$W&   0$''$N&   5.44 &$\pm$  0.34 \\
 & 10.5$''$&  12$''$W&  12$''$N&   3.79 \\
vdB  133 &  6.2$''$&  20$''$E&  30$''$N& $<$  4.70 \\
 &  6.2$''$&  30$''$E&  20$''$N& $<$  4.68 \\
 &  8.2$''$&  30$''$W&   0$''$N& $<$  4.51 \\
 &  8.2$''$&  30$''$E&   0$''$N& $<$  3.95 \\
 & 10.5$''$&  20$''$E&  30$''$N&   1.82 &$\pm$  0.42 \\
 & 10.5$''$&  30$''$E&  20$''$N&   1.07 &$\pm$  0.32 \\
\end{tabular}
\end{center}
\begin{center}
\begin{tabular}{lrrrrr}
\multicolumn{6}{c}{{\bf TABLE 3b, Continued}}\\[12pt]
\multicolumn{6}{c}{{\bf 2.2 $\mu$m Surface Brightness of Reflection
Nebulae}}\\[12pt]
&&\multicolumn{2}{c}{Nebular Offset$^b$}&$S _ \nu$ (2.2 $\mu$m)\\
Name&Ap$^a$&$\Delta \alpha$&$\Delta \delta$&(MJy sr$ ^ {-1}$)\\[12pt]
vdB  139 &  6.2$''$&  52$''$W&  34$''$N&   3.89 \\
 &  6.2$''$&  44$''$W&  27$''$N&   7.58 \\
 &  6.2$''$&  35$''$W&  20$''$N&   8.36 \\
 &  6.2$''$&  30$''$W&  20$''$N&   7.19 \\
 &  6.2$''$&   0$''$E&  30$''$N&   5.90 &$\pm$  0.34 \\
 & 10.5$''$&  30$''$W&  20$''$N&   9.19 \\
 & 10.5$''$&   0$''$E&  60$''$S&   1.89 &$\pm$  0.40 \\
 & 10.5$''$&   0$''$E&  30$''$N&   7.37 &$\pm$  0.41 \\
 & 10.5$''$&   0$''$E&  60$''$N&   2.70 &$\pm$  0.56 \\
 & 10.5$''$&   0$''$E& 120$''$N&   0.35 &$\pm$  0.10 \\
vdB  146 &  8.2$''$&  30$''$W&   0$''$N& $<$  2.44 \\
 &  8.2$''$&  30$''$E&   0$''$N&   5.25 &$\pm$  0.86 \\
vdB  152 &  8.2$''$&  30$''$W&   0$''$N& $<$  0.75 \\
 &  8.2$''$&  30$''$E&   0$''$N& $<$  1.42 \\
\end{tabular}
\end{center}

\noindent
($a$) Aperture diameter in arcseconds.

\noindent
($b$) Offset from central star of nebular position, in arcseconds.

\noindent
Note to Table---Uncertainties are $\pm$1-$\sigma$.  Upper limits are
3-$\sigma$.

\clearpage
\begin{center}
\begin{tabular}{lrrrrrrrr}
\multicolumn{9}{c}{{\bf TABLE 4}}\\[12pt]
\multicolumn{9}{c}{{\bf Infrared Colors of Reflection Nebulae}}\\[12pt]
&&\multicolumn{2}{c}{Nebular Offset$^b$}\\
Name&Ap$^a$&$\Delta \alpha$&$\Delta \delta$&$J-K$&$J-H$&$H-K$&$K-$[3.3
$\mu$m]&$K-L'$\\[12pt]
vdB   10 & 10.5$''$&  30$''$W&  30$''$N& $  0.70$ & $  0.41$ & $  0.29$ & $<
1.67$ & $<  0.42$ \\
&&&&$\pm$  0.12 &$\pm$  0.12 &$\pm$  0.11 & & \\
vdB   17 &  6.2$''$&   0$''$E&  20$''$S& $  1.72$ & $  0.71$ & $  1.01$ & $
4.39$ & $  3.39$ \\
&&&&$\pm$  0.22 &$\pm$  0.23 &$\pm$  0.20 &$\pm$  0.30 &$\pm$  0.32 \\
 &  8.2$''$&  30$''$E&  30$''$S&...&...& $>  0.81$ &...&...\\
&&&& & & & & \\
 & 10.5$''$&   0$''$E&  20$''$S& $  1.72$ & $  0.88$ & $  0.83$ & $  4.46$ & $
2.56$ \\
&&&&$\pm$  0.11 &$\pm$  0.12 &$\pm$  0.10 &$\pm$  0.11 &$\pm$  0.16 \\
vdB   22 & 10.5$''$&  30$''$W&  30$''$S& $>  0.87$ &...& $>  0.78$ & $<  4.21$
& $  2.93$ \\
&&&& & & & &$\pm$  0.29 \\
 & 10.5$''$&   0$''$E& 125$''$S& $> -0.29$ &...& $>  0.36$ &...&...\\
&&&& & & & & \\
 & 19.6$''$&  30$''$W&  30$''$S&...&...&...& $<  2.28$ & $<  1.77$ \\
&&&& & & & & \\
vdB   34 & 10.5$''$&  30$''$E&  30$''$S& $  0.22$ & $  0.30$ & $ -0.09$ & $<
3.02$ & $<  2.09$ \\
&&&&$\pm$  0.13 &$\pm$  0.14 &$\pm$  0.16 & & \\
 & 19.6$''$&  30$''$E&  30$''$S&...&...&...& $<  3.23$ & $<  2.56$ \\
&&&& & & & & \\
vdB   46 & 10.5$''$&  30$''$W&  30$''$N& $  0.38$ & $  0.35$ & $  0.03$
&...&...\\
&&&&$\pm$  0.17 &$\pm$  0.19 &$\pm$  0.19 & & \\
vdB   52 & 10.5$''$&  40$''$W&  40$''$S& $  1.83$ & $  0.94$ & $  0.89$ &...& $
 2.39$ \\
&&&&$\pm$  0.19 &$\pm$  0.16 &$\pm$  0.19 & &$\pm$  0.21 \\
 & 10.5$''$&   0$''$E&  60$''$S& $  2.13$ & $  1.25$ & $  0.88$ & $  3.87$ & $
2.30$ \\
&&&&$\pm$  0.19 &$\pm$  0.16 &$\pm$  0.19 &$\pm$  0.21 &$\pm$  0.21 \\
 & 10.5$''$&   0$''$E&  60$''$N& $  1.95$ & $  0.88$ & $  1.07$ & $  3.03$ & $
1.94$ \\
&&&&$\pm$  0.20 &$\pm$  0.21 &$\pm$  0.18 &$\pm$  0.26 &$\pm$  0.27 \\
vdB   57 & 10.5$''$&  30$''$E&  30$''$S& $  1.91$ & $  0.89$ & $  1.02$ & $
3.68$ & $<  2.10$ \\
&&&&$\pm$  0.18 &$\pm$  0.17 &$\pm$  0.16 &$\pm$  0.26 & \\
\end{tabular}
\end{center}
\begin{center}
\begin{tabular}{lrrrrrrrr}
\multicolumn{9}{c}{{\bf TABLE 4, Continued}}\\[12pt]
\multicolumn{9}{c}{{\bf Infrared Colors of Reflection Nebulae}}\\[12pt]
&&\multicolumn{2}{c}{Nebular Offset$^b$}\\
Name&Ap$^a$&$\Delta \alpha$&$\Delta \delta$&$J-K$&$J-H$&$H-K$&$K-$[3.3
$\mu$m]&$K-L'$\\[12pt]
vdB   59 & 10.5$''$&  60$''$E&  40$''$S& $  1.78$ & $  0.71$ & $  1.06$ & $
3.76$ & $  2.38$ \\
&&&&$\pm$  0.24 &$\pm$  0.31 &$\pm$  0.22 &$\pm$  0.25 &$\pm$  0.30 \\
vdB   60 & 10.5$''$&  30$''$E&  30$''$S& $  2.54$ & $  1.15$ & $  1.39$ & $
3.24$ & $  2.49$ \\
&&&&$\pm$  0.10 &$\pm$  0.11 &$\pm$  0.09 &$\pm$  0.13 &$\pm$  0.13 \\
vdB   72 & 10.5$''$&  30$''$W&  30$''$S& $  2.07$ & $  1.18$ & $  0.89$ & $
3.61$ & $  2.45$ \\
&&&&$\pm$  0.12 &$\pm$  0.11 &$\pm$  0.08 &$\pm$  0.20 &$\pm$  0.15 \\
vdB   74 & 10.5$''$&  30$''$E&  30$''$N& $  1.77$ & $  1.01$ & $  0.76$ & $<
3.40$ & $<  1.91$ \\
&&&&$\pm$  0.27 &$\pm$  0.33 &$\pm$  0.23 & & \\
 & 19.6$''$&  30$''$E&  30$''$N&...&...&...& $  1.79$ &...\\
&&&& & & &$\pm$  0.37 & \\
Parsamyan 18 &  6.2$''$&  12$''$W&   0$''$N&...&...&...& $  4.22$ & $  2.78$ \\
&&&& & & &$\pm$  0.21 &$\pm$  0.29 \\
 &  6.2$''$&  12$''$W&  12$''$N&...&...&...& $  4.39$ & $  2.82$ \\
&&&& & & &$\pm$  0.19 &$\pm$  0.30 \\
 &  6.2$''$&   0$''$E&  12$''$S&...&...&...& $  3.20$ & $  1.98$ \\
&&&& & & &$\pm$  0.20 &$\pm$  0.28 \\
 &  6.2$''$&   0$''$E&  12$''$N&...&...&...& $  4.42$ & $  2.70$ \\
&&&& & & &$\pm$  0.19 &$\pm$  0.28 \\
 &  6.2$''$&  12$''$E&   0$''$N&...&...&...& $  3.99$ & $  2.97$ \\
&&&& & & &$\pm$  0.27 &$\pm$  0.33 \\
 & 10.5$''$&  15$''$W&   0$''$N&...&...&...& $  4.12$ & $  2.98$ \\
&&&& & & &$\pm$  0.13 &$\pm$  0.10 \\
 & 10.5$''$&  12$''$W&  12$''$N& $  1.88$ & $  0.97$ & $  0.91$ & $  4.48$ & $
2.78$ \\
&&&&$\pm$  0.09 &$\pm$  0.10 &$\pm$  0.07 &$\pm$  0.13 &$\pm$  0.19 \\
vdB  133 & 10.5$''$&  20$''$E&  30$''$N& $  1.16$ & $  0.82$ & $  0.34$ &...&
$<  0.98$ \\
&&&&$\pm$  0.39 &$\pm$  0.39 &$\pm$  0.36 & & \\
 & 10.5$''$&  30$''$E&  30$''$N&...&...&...&...& $  0.89$ \\
&&&& & & & &$\pm$  0.31 \\
 & 19.6$''$&  30$''$E&  30$''$N&...&...&...& $<  1.38$ &...\\
&&&& & & & & \\
\end{tabular}
\end{center}
\begin{center}
\begin{tabular}{lrrrrrrrr}
\multicolumn{9}{c}{{\bf TABLE 4, Continued}}\\[12pt]
\multicolumn{9}{c}{{\bf Infrared Colors of Reflection Nebulae}}\\[12pt]
&&\multicolumn{2}{c}{Nebular Offset$^b$}\\
Name&Ap$^a$&$\Delta \alpha$&$\Delta \delta$&$J-K$&$J-H$&$H-K$&$K-$[3.3
$\mu$m]&$K-L'$\\[12pt]
vdB  139 &  6.2$''$&  52$''$W&  34$''$N&...&...&...& $  3.74$ & $  2.01$ \\
&&&& & & &$\pm$  0.13 &$\pm$  0.11 \\
 &  6.2$''$&  44$''$W&  27$''$N&...&...&...& $  4.47$ & $  2.63$ \\
&&&& & & &$\pm$  0.08 &$\pm$  0.08 \\
 &  6.2$''$&  35$''$W&  20$''$N&...&...&...& $  3.83$ & $  2.58$ \\
&&&& & & &$\pm$  0.09 &$\pm$  0.07 \\
 & 10.5$''$&  30$''$W&  20$''$N& $  2.04$ & $  1.11$ & $  0.93$ & $  3.69$ & $
2.24$ \\
&&&&$\pm$  0.07 &$\pm$  0.12 &$\pm$  0.12 &$\pm$  0.16 &$\pm$  0.08 \\
 & 10.5$''$&   0$''$E&  60$''$S&...&...&...& $  4.03$ & $  2.55$ \\
&&&& & & &$\pm$  0.32 &$\pm$  0.35 \\
 & 10.5$''$&   0$''$E&  30$''$N& $  2.10$ & $  1.08$ & $  1.02$ & $  3.60$ & $
2.31$ \\
&&&&$\pm$  0.10 &$\pm$  0.10 &$\pm$  0.09 &$\pm$  0.22 &$\pm$  0.10 \\
 & 10.5$''$&   0$''$E&  60$''$N&...&...&...& $  3.89$ & $  2.20$ \\
&&&& & & &$\pm$  0.32 &$\pm$  0.33 \\
 & 10.5$''$&   0$''$E& 120$''$N&...&...&...& $<  4.92$ & $<  3.80$ \\
&&&& & & & & \\
vdB  146 &  6.2$''$&  30$''$E&  30$''$N& $  1.10$ & $  0.60$ & $  0.50$ &...& $
 2.72$ \\
&&&&$\pm$  0.16 &$\pm$  0.17 &$\pm$  0.14 & &$\pm$  0.27 \\
 &  9.2$''$&  30$''$E&  30$''$N&...&...&...& $  4.38$ & $  3.44$ \\
&&&& & & &$\pm$  0.20 &$\pm$  0.21 \\
\end{tabular}
\end{center}
\noindent
($a$) Aperture diameter in arcseconds.

\noindent
($b$) Offset from central star of nebular position, in arcseconds.

\noindent
Note to Table---Uncertainties are $\pm$1-$\sigma$.
Upper and lower limits are 3-$\sigma$.

\clearpage
\begin{center}
\begin{tabular}{lrrrrrr}
\multicolumn{7}{c}{{\bf TABLE 5}}\\[12pt]
\multicolumn{7}{c}{{\bf Infrared Magnitudes of Illuminating Stars of
Reflection Nebulae}}\\[12pt]
\multicolumn{1}{l}{Name}&\multicolumn{1}{r}{Ap$^a$}&\multicolumn{1}{r}{$J$}
&\multicolumn{1}{r}{$H$}&\multicolumn{1}{r}{$K$}
&\multicolumn{1}{r}{[3.3 $\mu$m]}&\multicolumn{1}{r}{$L'$}\\[12pt]
vdB   10 &  7.7$''$&...&...&   3.75 &...&...\\
&& & &$\pm$  0.10 & & \\
 & 10.5$''$&...&   3.96 &   3.85 &   3.65 &   3.46 \\
&& & & &$\pm$  0.16 &$\pm$  0.10 \\
vdB   16 & 10.5$''$&...&...&   7.64 &...&...\\
&& & &$\pm$  0.06 & & \\
vdB   17 &  6.2$''$&   8.81 &   8.49 &   8.30 &   8.06 &   8.06 \\
&&$\pm$  0.15 &$\pm$  0.11 &$\pm$  0.11 &$\pm$  0.14 &$\pm$  0.24 \\
 &  8.2$''$&   8.87 &   8.52 &   8.30 &...&...\\
&&$\pm$  0.07 &$\pm$  0.13 &$\pm$  0.18 & & \\
 & 10.5$''$&   8.84 &   8.49 &   8.31 &   7.82 &   8.10 \\
&&$\pm$  0.06 & & & & \\
vdB   22 &  8.2$''$&...&...&   4.17 &...&...\\
&& & &$\pm$  0.18 & & \\
 & 10.5$''$&...&...&   4.28 &   4.29 &   4.26 \\
&& & & &$\pm$  0.09 & \\
 & 19.6$''$&...&...&   4.32 &   4.32 &   4.49 \\
&& & &$\pm$  0.10 &$\pm$  0.21 &$\pm$  0.20 \\
 & 57.0$''$&   4.29 &   4.24 &   4.24 &...&...\\
&& & & & & \\
Elias 1 &  8.2$''$&...&...&   5.60 &...&...\\
&& & &$\pm$  0.18 & & \\
 & 10.5$''$&   8.78 &   6.95 &   5.80 &   4.72 &   4.41 \\
&&$\pm$  0.07 & &$\pm$  0.06 &$\pm$  0.10 &$\pm$  0.10 \\
vdB   34 &  6.2$''$&   5.32 &   5.31 &   5.27 &   5.25 &   5.22 \\
&&$\pm$  0.15 &$\pm$  0.11 &$\pm$  0.11 &$\pm$  0.14 &$\pm$  0.24 \\
 & 19.6$''$&...&...&   5.23 &   5.34 &   5.49 \\
&& & &$\pm$  0.10 &$\pm$  0.21 &$\pm$  0.20 \\
vdB   35 & 10.5$''$&...&...&   6.12 &...&...\\
&& & &$\pm$  0.19 & & \\
\end{tabular}
\end{center}
\begin{center}
\begin{tabular}{lrrrrrr}
\multicolumn{7}{c}{{\bf TABLE 5, Continued}}\\[12pt]
\multicolumn{7}{c}{{\bf Infrared Magnitudes of Illuminating Stars of
Reflection Nebulae}}\\[12pt]
\multicolumn{1}{l}{Name}&\multicolumn{1}{r}{Ap$^a$}&\multicolumn{1}{r}{$J$}
&\multicolumn{1}{r}{$H$}&\multicolumn{1}{r}{$K$}
&\multicolumn{1}{r}{[3.3 $\mu$m]}&\multicolumn{1}{r}{$L'$}\\[12pt]
vdB   46 & 10.5$''$&   8.19 &   7.07 &   6.20 &...&...\\
&&$\pm$  0.07 & &$\pm$  0.06 & & \\
vdB   52 &  6.2$''$&...&...&   7.37 &...&   7.01 \\
&& & &$\pm$  0.15 & &$\pm$  0.11 \\
 & 10.5$''$&   7.55 &   7.45 &   7.38 &   6.85 &   7.10 \\
&&$\pm$  0.11 &$\pm$  0.11 &$\pm$  0.12 &$\pm$  0.21 &$\pm$  0.10 \\
vdB   57 & 10.5$''$&   7.64 &   7.37 &   7.25 &   7.32 &   7.21 \\
&& & & & & \\
vdB   59 & 10.5$''$&   7.64 &   7.13 &   6.82 &   6.63 &   6.73 \\
&&$\pm$  0.21 &$\pm$  0.21 &$\pm$  0.21 &$\pm$  0.21 &$\pm$  0.21 \\
vdB   60 &  8.2$''$&   7.05 &...&   6.81 &...&...\\
&&$\pm$  0.07 & &$\pm$  0.19 & & \\
 & 10.5$''$&   7.11 &   6.60 &   6.42 &   6.27 &   6.16 \\
&&$\pm$  0.07 & &$\pm$  0.06 &$\pm$  0.10 &$\pm$  0.10 \\
vdB   72 &  6.2$''$&...&...&   8.65 &...&...\\
&& & & & & \\
 & 10.5$''$&   8.71 &   8.63 &   8.59 &...&   8.31 \\
&& & & & &$\pm$  0.07 \\
vdB   74 & 10.5$''$&   9.70 &   9.48 &   9.25 &   8.23 &   8.43 \\
&&$\pm$  0.18 &$\pm$  0.18 &$\pm$  0.06 &$\pm$  0.10 &$\pm$  0.10 \\
Parsamyan 18 &  6.2$''$&  10.32 &   9.68 &   9.19 &   7.46 &   7.92 \\
&&$\pm$  0.15 &$\pm$  0.11 &$\pm$  0.11 &$\pm$  0.14 &$\pm$  0.24 \\
 & 10.5$''$&...&...&...&   6.74 &...\\
&& & & &$\pm$  0.23 & \\
vdB  101 &  7.7$''$&...&...&   3.67 &...&...\\
&& & &$\pm$  0.10 & & \\
vdB  111 &  7.7$''$&...&...&   5.60 &...&...\\
&& & & & & \\
\end{tabular}
\end{center}
\begin{center}
\begin{tabular}{lrrrrrr}
\multicolumn{7}{c}{{\bf TABLE 5, Continued}}\\[12pt]
\multicolumn{7}{c}{{\bf Infrared Magnitudes of Illuminating Stars of
Reflection Nebulae}}\\[12pt]
\multicolumn{1}{l}{Name}&\multicolumn{1}{r}{Ap$^a$}&\multicolumn{1}{r}{$J$}
&\multicolumn{1}{r}{$H$}&\multicolumn{1}{r}{$K$}
&\multicolumn{1}{r}{[3.3 $\mu$m]}&\multicolumn{1}{r}{$L'$}\\[12pt]
vdB  133 &  6.2$''$&...&...&...&...&   3.12 \\
&& & & & &$\pm$  0.13 \\
 &  8.2$''$&   4.24 &   3.78 &   3.58 &...&...\\
&&$\pm$  0.10 &$\pm$  0.13 &$\pm$  0.18 & & \\
 & 10.5$''$&...&...&...&...&   3.44 \\
&& & & & &$\pm$  0.08 \\
 & 19.6$''$&...&...&...&   3.56 &...\\
&& & & &$\pm$  0.21 & \\
vdB  135 &  7.7$''$&...&...&   3.57 &...&...\\
&& & &$\pm$  0.10 & & \\
vdB  137 & 10.5$''$&...&   4.45 &   4.35 &...&...\\
&& & & & & \\
vdB  139 &  6.2$''$&...&...&   4.76 &...&   3.36 \\
&& & & & & \\
 & 10.5$''$&   6.09 &   5.41 &   4.63 &   3.58 &   3.37 \\
&& & & &$\pm$  0.21 & \\
vdB  146 &  6.2$''$&...&...&   8.48 &...&   8.00 \\
&& & & & &$\pm$  0.06 \\
 &  8.2$''$&...&...&   8.34 &...&...\\
&& & &$\pm$  0.11 & & \\
 &  9.2$''$&...&...&   8.44 &   8.02 &   8.21 \\
&& & & &$\pm$  0.12 &$\pm$  0.09 \\
 & 10.5$''$&   8.83 &   8.57 &   8.36 &...&...\\
&&$\pm$  0.07 & &$\pm$  0.06 & & \\
vdB  152 &  8.2$''$&   8.93 &   8.85 &   9.04 &...&...\\
&&$\pm$  0.07 &$\pm$  0.13 &$\pm$  0.11 & & \\
\end{tabular}
\end{center}
\noindent
($a$) Aperture diameter in arcseconds.

\noindent
Note to Table---Uncertainties are $\pm$1-$\sigma$,
and are only given when larger than 0.05 mag.

\clearpage

\begin{center}
\begin{tabular}{lcccc}
\multicolumn{3}{c}{{\bf TABLE 6}}\\[12pt]
\multicolumn{3}{c}{{\bf 10 and 20 Micron Magnitudes of}}\\
\multicolumn{3}{c}{{\bf Illuminating Stars of Reflection Nebulae}}\\[12pt]
Nebula&$N$&$Q$\\[12pt]
vdB 9&4.10 $\pm$ 0.10&...\\
vdB 10&3.63 $\pm$ 0.08&...\\
vdB 12&3.53&3.35 $\pm$ 0.09\\
vdB 16&7.64 $\pm$ 0.15&...\\
17 Tau&3.77&3.72 $\pm$ 0.14\\
20 Tau&3.99 $\pm$ 0.09&...\\
23 Tau&3.92 $\pm$ 0.08&3.70 $\pm$ 0.19\\
25 Tau&2.43&2.04 $\pm$ 0.08\\
vdB 25&4.47&$>$4.46\\
AE Aur&5.37 $\pm$ 0.11&...\\
vdB 35&5.98 $\pm$ 0.07&...\\
vdB 37&0.84&0.69\\
vdB 40&5.78 $\pm$ 0.08&...\\
vdB 41&8.13 $\pm$ 0.34&...\\
vdB 47&2.68 $\pm$ 0.08&2.66 $\pm$ 0.08\\
NGC 2068&5.66 $\pm$ 0.10&...\\
NGC 2071&6.51 $\pm$ 0.24&...\\
vdB 61&5.59 $\pm$ 0.10&...\\
vdB 62&6.40 $\pm$ 0.12&...\\
vdB 133&3.46 $\pm$ 0.10&3.50 $\pm$ 0.26\\
vdB 136&3.13 $\pm$ 0.10&2.88 $\pm$ 0.24\\
vdB 135&3.39 $\pm$ 0.10&3.53 $\pm$ 0.28\\
vdB 137&4.07 $\pm$ 0.10&...\\[12pt]
\end{tabular}
\end{center}

\noindent
Note to Table---Uncertainties are $\pm$1-$\sigma$, and are only given when
larger than 0.05 mag.  Lower limits are 3-$\sigma$.
Aperture diameter was 6$''$ for all measurements.

\clearpage
\begin{center}
\begin{tabular}{lll}
\multicolumn{3}{c}{{\bf TABLE 7}}\\[12pt]
\multicolumn{3}{c}{{\bf Estimated 2.2 $\mu$m Nebular Optical Depths}}\\[12pt]
Nebula  &$\tau(K)^a$ & $\tau(K)^b$\\
   10 & ... &   0.0027 \\
   17 &    0.098 --   0.155 &    0.108 \\
   22 &   0.0066 --  0.0843 &   0.0140 \\
   34 & ... &  0.00051 \\
   35 & ... &    0.030 \\
   46 &     0.18 --    0.57 & ... \\
   52 &    0.083 --   0.123 &    0.108 \\
   57 &   0.0170 &   0.0073 \\
   59 &     1.32 --    2.48 &     0.11 \\
   60 &    0.060 --   0.275 &    0.108 \\
   72 &    0.068 &    0.108 \\
   74 &    0.021 --   0.038 &    0.108 \\
 9918 &     0.16 --    1.08 &     0.11 \\
  101 & ... &   0.0067 \\
  111 & ... &   0.0075 \\
  133 & ... &   0.0080 \\
  135 & ... &    0.028 \\
  137 &  0.00095 & ... \\
  139 &    0.045 --   0.093 &    0.055 \\
  146 &    0.066 --   0.166 & ... \\
  152 &    0.014 --   0.079 & ... \\
\end{tabular}
\end{center}

Notes to Table:

($a$) Nebular optical depth at $K$, $\tau$($K$),
derived from Equation 3 combined with visual observations
of reflected light
(Racine 1971;
Witt 1977;
Witt \& Cottrell 1980$a$;
Cottrell 1981;
Witt, Schild, \& Kraiman 1984;
Witt 1985$b$, 1986;
Witt \& Schild 1986;
Witt et al. 1987)
to estimate the visual optical depth, $\tau$($V$), for each nebula,
with $\tau$($K$) = $\tau$($V$) / 9.29
(Mathis 1990).

($b$) Nebular optical depth at $K$, $\tau$($K$),
derived from the value of $A_V(neb)$ calculated in Paper III,
which is an estimate of
$\tau$($V$) for each nebula calculated from the
amount of starlight incident on the nebula which is
reradiated in the far infrared,
with $\tau$($K$) = $\tau$($V$) / 9.29
(Mathis 1990).

\clearpage

\centerline
{\bf Figure Captions}

\vskip 12pt
{\bf Figure 1}--- Broadband energy distribution {\it (filled circles)}
of nebular surface brightness
from NGC 2071, 30$''$E 30$''$S of the central star,
together with the spectrum near
the 3.3 $\mu$m emission feature {\it (histogram)}.
The 3.2--3.6 $\mu$m spectrum was measured with a
circular variable filter with a spectral resolution
$\lambda$/$\Delta \lambda$ = 67 and
a 10.5$''$ diameter beam.
Units are $\nu S _ \nu$
in W m$^{-2}$ sr$^{-1}$, where $\nu$ is the frequency
and $S _ \nu$ is the surface brightness at that frequency,
and wavelength in $\mu$m.
Uncertainties are $\pm$1-$\sigma$.
The data at $J$, $H$, $K$, $L'$, and near the 3.3 $\mu$m feature
are from this paper.
Visual data are from Witt \& Schild (1986).
An estimate of the surface brightness of scattered starlight
(see text),
normalized at $V$, is shown assuming either
{\it (solid line)}
the albedo is independent of wavelength
or {\it (dotted line)}
the albedo depends
on wavelength as predicted by Draine \& Lee (1984).

{\bf Figure 2}--- Near infrared colors $J-H$ and $H-K$ of extended emission
in reflection nebulae.
Nebulae
in which the 3.3 $\mu$m emission feature was detected ({\it filled circles}),
are plotted separately
from nebulae
({\it open circles})
in which the 3.3 $\mu$m emission feature
was not searched
for, or searched for and not detected.
The near infrared colors of the
central stars with ({\it filled stars})
and without ({\it open stars}) hydrogen emission
lines are also shown.
Uncertainties shown are $\pm$1-$\sigma$; no uncertainties
are shown for the stellar colors.
No limits on nebular colors are
shown but they are consistent with the other nebular colors shown.

{\bf Figure 3}--- $V-K$ color of extended emission
in reflection nebulae, $(V-K)_{\rm neb\ }$,
plotted vs. the
$V-K$ color of the illuminating star, $(V-K)_{{\rm star\ }}$.
Nebulae
in which the 3.3 $\mu$m emission feature was detected ({\it filled circles}),
are plotted separately
from nebulae
({\it open circles})
in which the 3.3 $\mu$m emission feature
was not searched
for, or searched for and not detected.
The $V-K$ colors of the purely scattered light component in
NGC 7023 (Paper IV) are also shown ({\it open triangles}).
Upper limits are 3-$\sigma$; uncertainties shown
are $\pm$1-$\sigma$.
$(V-K)_{\rm neb}$
as a function of $(V-K)_{\rm star\ }$, predicted for scattered starlight
(see text),
is shown for four cases with different assumptions about
how the albedo depends
on wavelength and whether the nebula is optically thin or thick:
(1) optically thick nebula, high albedo grains {\it (dotted line)};
(2) optically thick nebula, low albedo grains {\it (short-dashed line)};
(3) optically thin nebula, high albedo grains {\it (long-dashed line)};
and (4) optically thin nebula, low albedo grains {\it (solid line)}.
Nebulae with $K$ emission in excess of scattered starlight
fall above and to the left of these lines.

{\bf Figure 4}--- Plot of the logarithm of
the ratio of the observed $K$ surface brightness, $S_{\rm obs\ }$,
to the estimated surface brightness of reflected starlight,
$S_{\rm ref}$(est), vs. log($T_{\rm star}$), where
$T_{\rm star}$ is the temperature of the central star of each nebula.
Nebulae
in which the 3.3 $\mu$m emission feature was detected ({\it filled circles}),
are plotted separately
from nebulae
({\it open circles})
in which the 3.3 $\mu$m emission feature
was not searched
for, or searched for and not detected.
Upper limits are 3-$\sigma$.
The purely scattered light component of
the $K$ emission of NGC 7023 (Paper IV) is also shown ({\it open triangles}).
The uncertainty in $S_{\rm ref}$(est) is a factor of 3, corresponding to
an uncertainty in log[$S_{\rm ref}$(est)] of 0.5.
Sources falling above the dotted line by more than this uncertainty
likely have excess $K$ emission
due to something other than reflected starlight.

{\bf Figure 5}---
The strength of the 3.3 $\mu$m emission feature in reflection nebulae
vs. temperature of the central star, $T_{\rm star\ }$.
Upper and lower limits are 3-$\sigma$; uncertainties shown
are $\pm$1-$\sigma$.
{\it Top:} $K$ magnitude minus the magnitude at
3.3 $\mu$m.
The 3.3 $\mu$m magnitude includes both 3.3 $\mu$m continuum emission
and any 3.3 $\mu$m feature emission.
{\it Bottom:} The surface brightness
(MJy sr$^{-1}$) at the peak of the 3.3 $\mu$m feature
ratioed to the continuum surface brightness (MJy sr$^{-1}$) at 3.3 $\mu$m.
The 3.3 $\mu$m feature emission has been corrected for
the continuum emission at 3.3 $\mu$m,
by interpolating between the $K$ and $L'$
surface brightnesses.

{\bf Figure 6}---
The surface brightness
(MJy sr$^{-1}$) at the peak of the 3.3 $\mu$m feature
ratioed to the continuum surface brightness (MJy sr$^{-1}$) at 3.3 $\mu$m,
vs. the $K-L'$ color.
The 3.3 $\mu$m feature emission has been corrected for
the continuum emission at 3.3 $\mu$m,
by interpolating between the $K$ and $L'$
surface brightnesses.
Upper and lower limits are 3-$\sigma$; uncertainties shown
are $\pm$1-$\sigma$.

{\bf Figure 7}---
Near infrared colors $J-H$, $H-K$ and $K-L'$ of extended emission
in reflection nebulae
plotted vs. temperature
of the central star, $T_{\rm star\ }$.
Nebulae
in which the 3.3 $\mu$m emission feature was detected ({\it filled circles}),
are plotted separately
from nebulae
({\it open circles})
in which the 3.3 $\mu$m emission feature
was not searched
for, or searched for and not detected.
Uncertainties shown are $\pm$1-$\sigma$.
The 3-$\sigma$ upper limits are given only for $K-L'$;
the lower limits to $J-H$ and $H-K$ are consistent with
the other plotted points.

\pagestyle{empty}

\clearpage

\begin{figure}
\plotfiddle{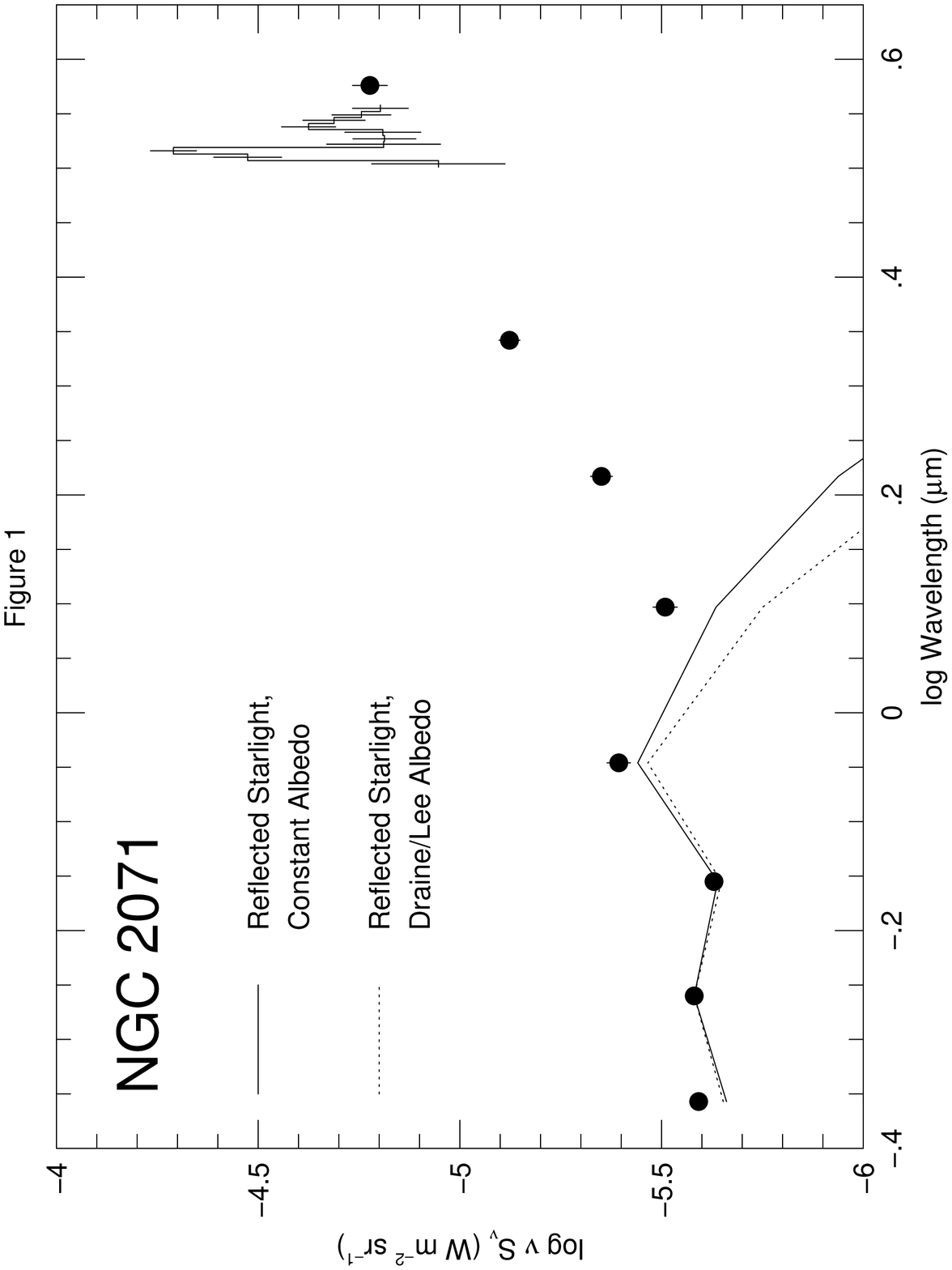}{9.0in}{90}{100}{100}{324}{0}
\end{figure}

\begin{figure}
\plotfiddle{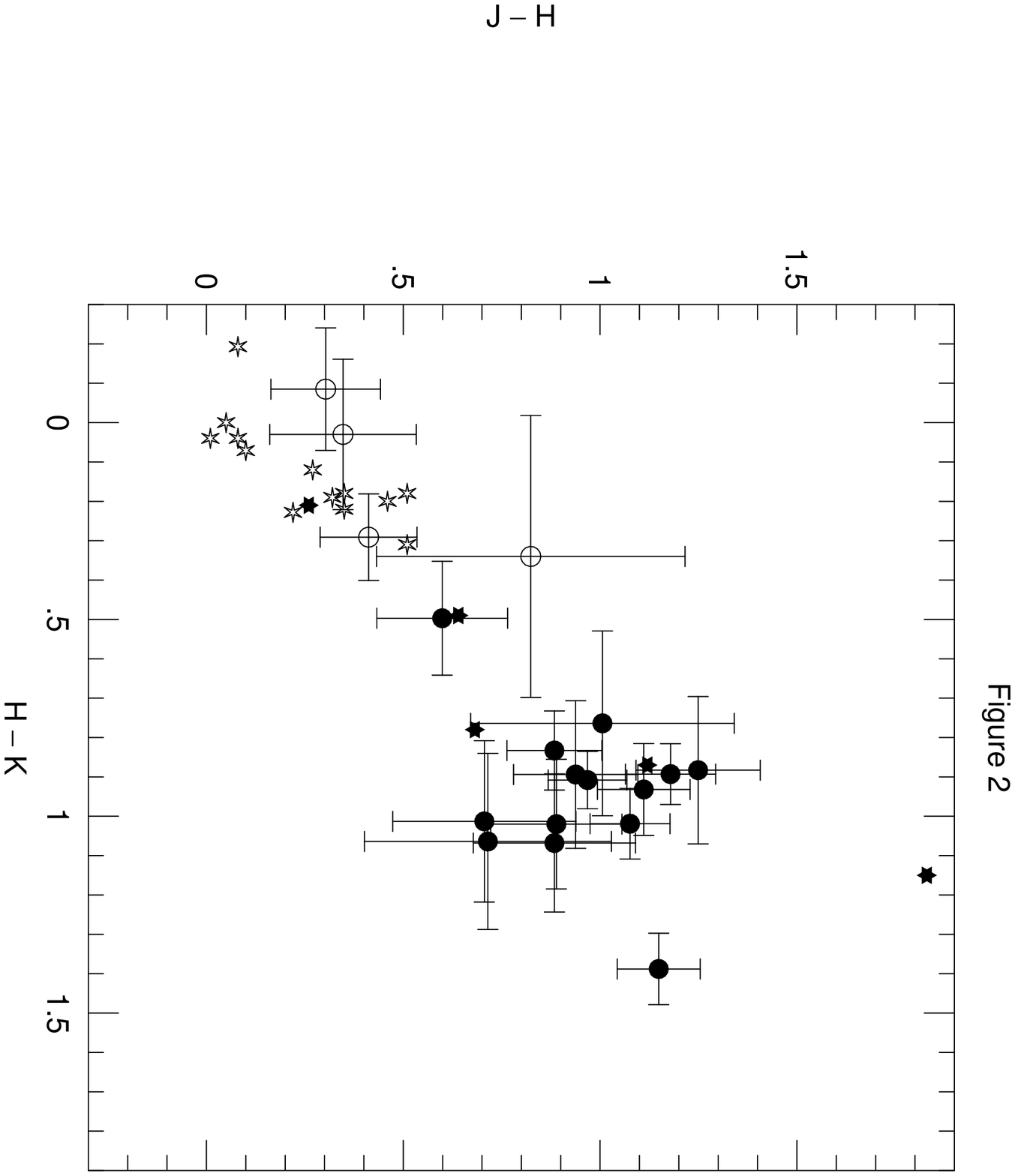}{9.0in}{90}{100}{100}{324}{0}
\end{figure}

\begin{figure}
\plotfiddle{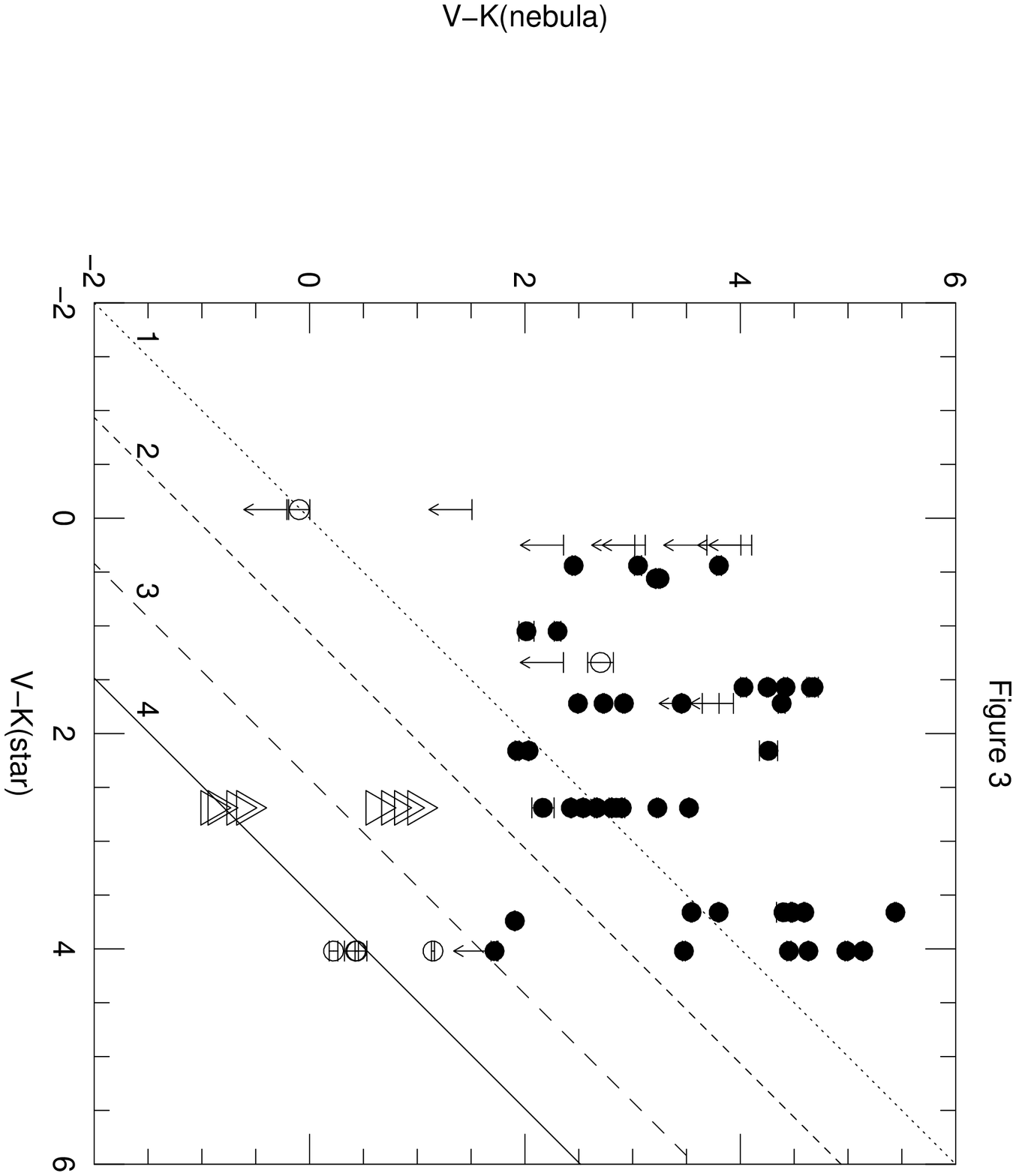}{9.0in}{90}{100}{100}{324}{0}
\end{figure}

\begin{figure}
\plotfiddle{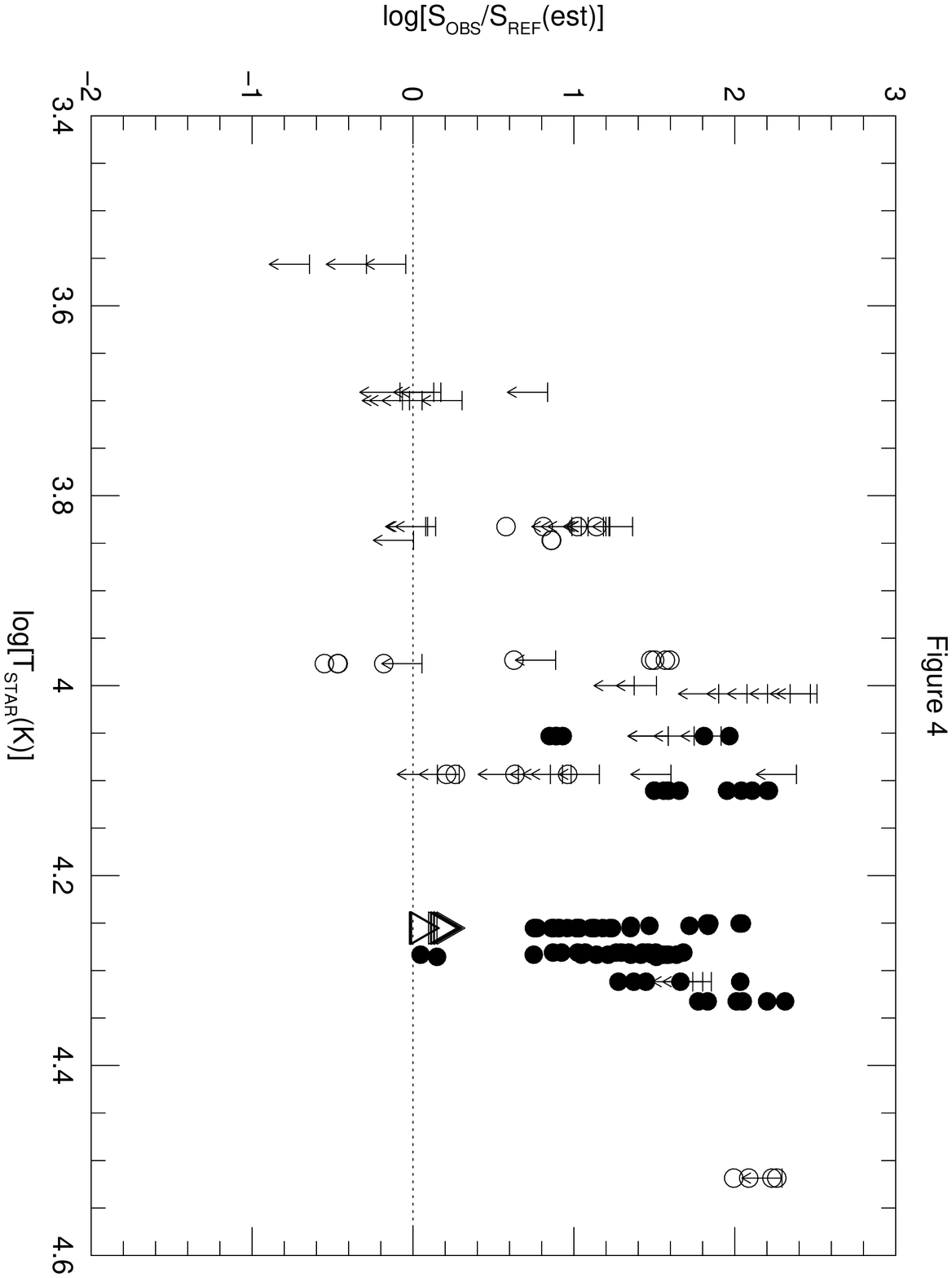}{9.0in}{90}{100}{100}{324}{0}
\end{figure}

\begin{figure}
\plotfiddle{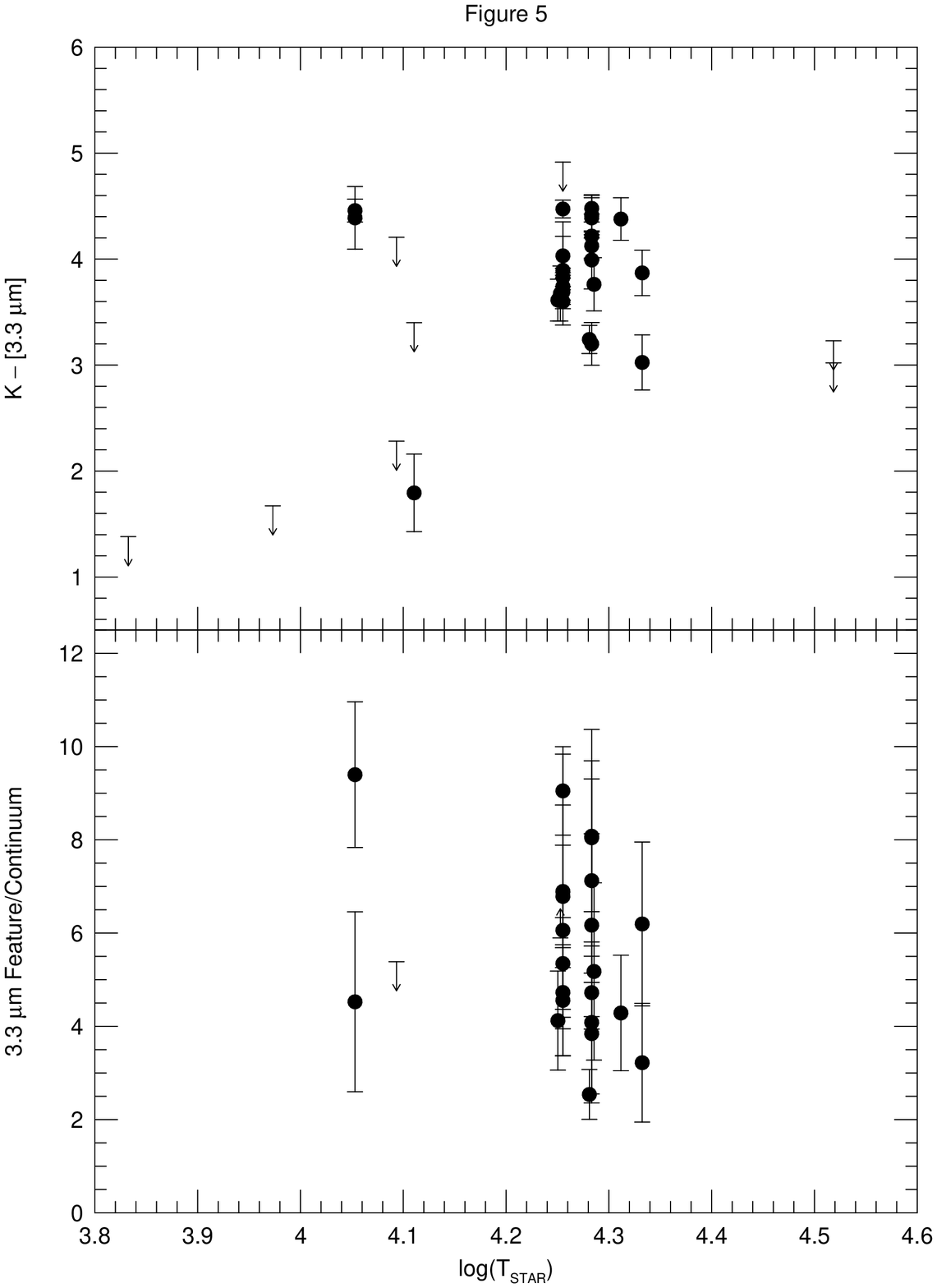}{9.0in}{0}{100}{100}{-324}{0}
\end{figure}

\begin{figure}
\plotfiddle{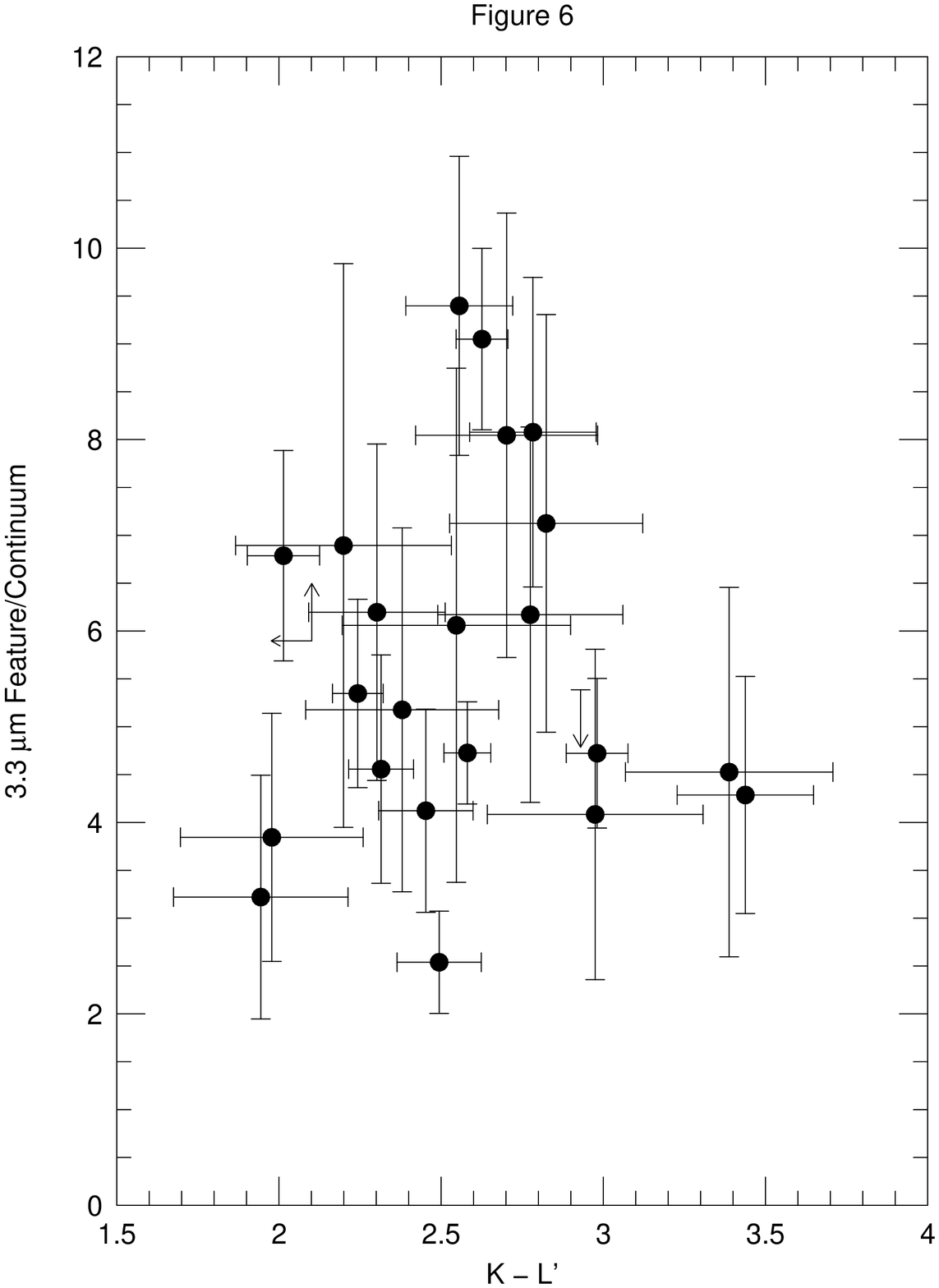}{9.0in}{90}{100}{100}{324}{0}
\end{figure}

\begin{figure}
\plotfiddle{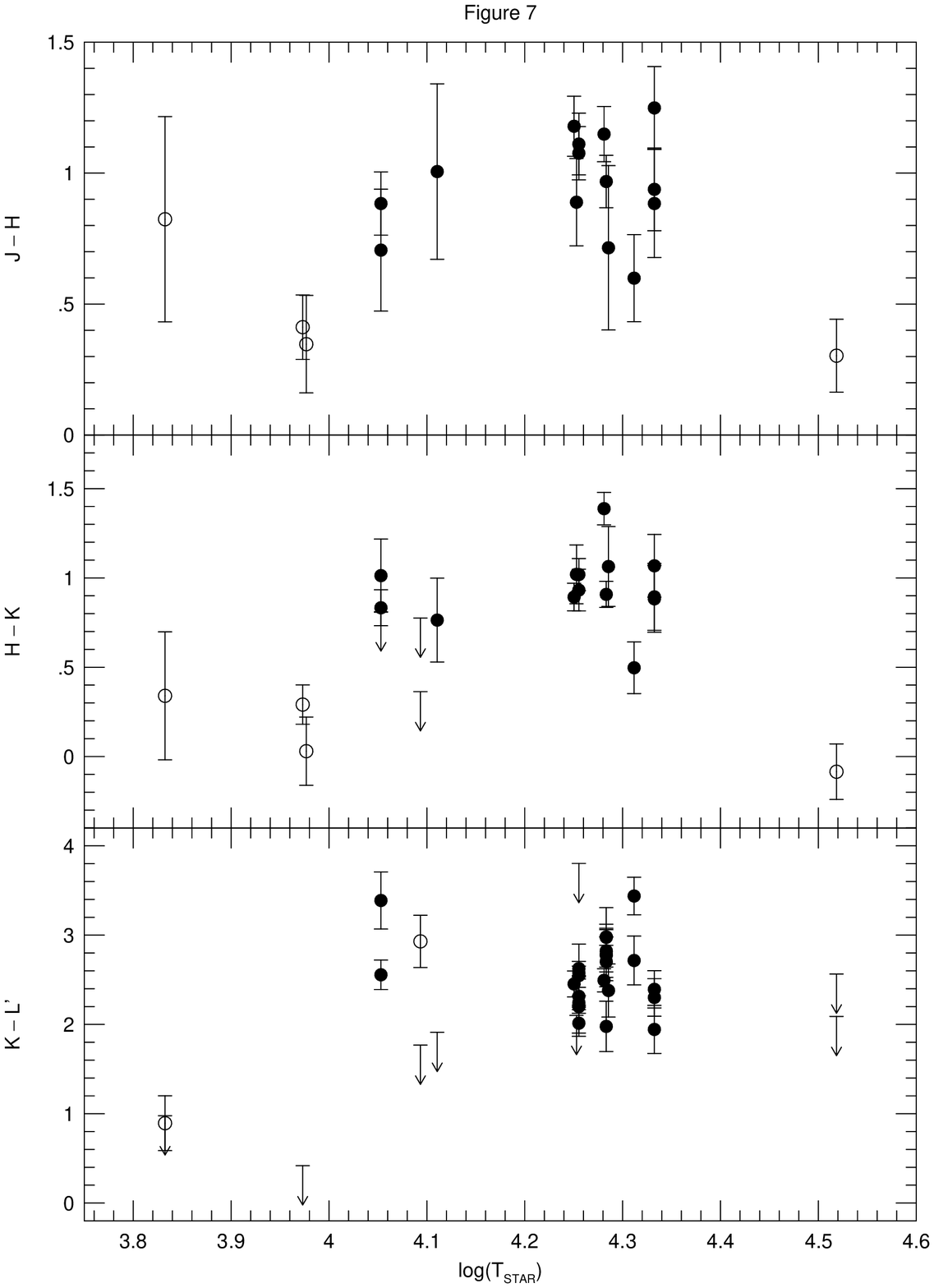}{9.0in}{0}{100}{100}{-324}{0}
\end{figure}

\end{document}